\documentclass[lettersize,journal]{IEEEtran}
\usepackage{amsmath,amsfonts}
\usepackage{algorithmic}
\usepackage{array}
\usepackage{textcomp}
\usepackage{stfloats}
\usepackage{url}
\usepackage{verbatim}
\usepackage{graphicx}
\def\BibTeX{{\rm B\kern-.05em{\sc i\kern-.025em b}\kern-.08em
    T\kern-.1667em\lower.7ex\hbox{E}\kern-.125emX}}
\usepackage{balance}

\usepackage[inline]{enumitem}
\usepackage{epstopdf}
\usepackage{booktabs} 
\usepackage[hidelinks]{hyperref}

\usepackage[dvipsnames]{xcolor}
\usepackage[deletedmarkup=sout,authormarkup=superscript]{changes}
\definechangesauthor[name={Samuel}, color=NavyBlue]{SB}
\definechangesauthor[name={Dominic}, color=RedViolet]{DLM}
\definechangesauthor[name={Alisa}, color=Orange]{AR}
\definechangesauthor[name={ToDo}, color=Green]{TODO}

\newcommand{\mc}[0] {\mathcal}

\usepackage[c1,short]{optidef}
\usepackage{subcaption}

\newcommand{\reals}{\mathbb{R}}
\newcommand{\OutputOne}{\gamma}
\newcommand{\OutputTwo}{\gamma}
\newcommand{\LocalVar}{\nu}
\newcommand{\ReferenceOne}{r}
\newcommand{\ReferenceTwo}{r}
\newcommand{\Objective}{\Xi}
\newcommand{\Bound}{\mu}
\newcommand{\DeltatOne}{\Delta \tau}
\newcommand{\DeltatTwo}{\Delta t}
\newcommand{\Transpose}{\top}
\newcommand{\Letters}{Letters}
\newcommand{\Airfoil}{airfoil}

\newcommand\resultsscaling{.78}

\begin{document}
\title{Data-driven Reference Trajectory Optimization for Precision Motion Systems}
\author{Samuel Balula, Dominic Liao-McPherson, Alisa Rupenyan, and John Lygeros
\thanks{
The authors are with the ETH Z\"{u}rich Automatic Control Laboratory, Physikstrasse 3, 8092 Z\"{u}rich, Switzerland. E-mail: \texttt{\{sbalula, dliaomc, ralisa, jlygeros\}@ethz.ch}. S. Balula and A. Rupenyan are also with Inspire AG.
This project has been funded by the Swiss Innovation Agency (Innosuisse, Grant Number 46716) and the Swiss National Science Foundation through NCCR Automation, a National Centre of Competence in Research (Grant Number 180545).}
}
\maketitle

\begin{abstract}
We propose a data-driven optimization-based pre-compensation method to improve the contour tracking performance of precision motion stages by modifying the reference trajectory and without modifying any built-in low-level controllers.
The position of the precision motion stage is predicted with data-driven models, a linear low-fidelity model is used to optimize traversal time, by changing the path velocity and acceleration profiles then a non-linear high-fidelity model is used to refine the previously found time-optimal solution.
We experimentally demonstrate that the proposed method is capable of simultaneously improving the productivity and accuracy of a high precision motion stage.
Given the data-based nature of the models, the proposed method can easily be adapted to a wide family of precision motion systems.

\end{abstract}


\section{Introduction} \label{sec:introduction}
A precision motion stage (PMS) is a robotic setup designed to precisely position an end-effector, typically in the micro- to nanometer range. PMS are used in lithography~\cite{kim1998high},
micro-assembly~\cite{gorman2003force},
precision metrology~\cite{kim20223},
and precision engineering~\cite{Khosravi2022}, 
and many other applications~\cite{tan2007precision,ouyang2008micro}.
They are often used as components in machine tools for the production of parts, such as metal laser cutting or grinding processes where the goal is to trace a desired target geometry as quickly and accurately as possible.

The accuracy of PMS systems is critical as the parts they produce must satisfy tight engineering tolerances. At the same time, economic considerations dictate rapid production of parts to boost productivity. Unfortunately, the physical inertia of PMS systems dictates a trade-off between productivity and accuracy and, in practice, the quality of the control system is critical for managing this trade-off. Typical industrial precision systems achieve high-precision using well-tuned ``classical'' feedback controllers, such as Proportional-Integral-Derivative (PID), to track a given target trajectory. These low-level controllers are designed by the manufacturer and users are typically not able to modify/tune the control algorithm.

In this paper, we propose an optimization-based reference (input trajectory) design (pre-compensation) methodology for simultaneously improving the precision and productivity of PMS by leveraging the widespread availability of system data. Our methodology is guided by the following requirements:
\begin{enumerate*}
	\item[(i)]   it does not require modifications to the existing hardware and low-level controllers, making it suitable for application in production machines;\label{guide1}
	\item[(ii)]  it does not require a physics-based model of the machine, relying exclusively on experimental data;\label{guide2}
	\item[(iii)] it can be applied to individual machines with diverse technical characteristics; and \label{guide3}
	\item[(iv)]  it can produce good results for previously unseen parts at the first attempt.\label{guide4}
\end{enumerate*}

These requirements are motivated by the requirements of ``Industry 4.0'' where machines are no longer standalone setups and are instead connected in a network, making data more easily available~\cite{ghobakhloo2018future}. This data can be exploited to improve performance; for example, it can be leveraged to create data-driven models, avoiding the labour intensive process of developing and maintaining first principles based ones~\cite{kouraytem2021modeling}. 
Future manufacturing is also expected to be more distributed and personalized (e.g., ``Manufacturing as a service'')~\cite{srai2016distributed}, leading to smaller volumes for any given part, possibly produced across a pool of non-homogeneous machines. This makes it more important to exploit data to ensure that machines can execute designs with minimal calibration. 

Several methods for improving PMS performance have been proposed in literature. 
One method is to re-designing the feedback controller. A common approach is model predictive contouring control (MPCC)~\cite{lam2010model}, 
wherein a model of contouring error is minimized in a receding horizon fashion, drawing on methods from autonomous racing ~\cite{Vazques2020}. However these methods have high online computational loads, rely on accurate process models, and require modifying the inner-loop controllers.

Another common approach in manufacturing and precision motion systems is {\it learning-based control}~\cite{wang2009survey, balta2021learning}. 
These methods assume that the processes is repeatable and adjust the input trajectories, using the error from the previous repetition as feedback. The input trajectories can be modified between iterations (Iterative Learning Control, and Run-to-Run control), or periods (Repetitive Control). These methods can exploit models~\cite{barton2010norm}, 
include constraints~\cite{liao2022robustness} 
, and are effective for repeatable processes. However, the computed trajectory is specific for a particular part and machine, with multiple trials required to achieve the desired accuracy. While this may be acceptable in large production runs, it may be inefficient for smaller runs e.g., in "manufacturing as a service" applications. 

The final major class of methods is pre-compensation, which 
aim to modify the input trajectory offline to reduce the contouring error for arbitrary parts~\cite{zhang2013pre}. A common approach is to drive a model of the contouring error to zero by manipulating the input trajectory using a classical controller (e.g., a PID)~\cite{lo1998cnc}. These methods cannot include look-ahead or constraints, motivating the use of offline MPCC algorithms based on physics-based nonlinear~\cite{haas2019mpcc}, identified linear models~\cite{yang2019model} or combined with a reference governor~\cite{di2018cascaded}. In~\cite{WangPrediction} the compensation is computed online using previous measurements and a linear model.
%
%
In~\cite{jiAng2022contour} the contour error is predicted with an Artificial Neural Network (ANN), and the pre-compensation computed with reinforcement learning. This approach does not allow constraints to be included and uses only zeroth-order information about the ANN model, despite the ready availability of ANN derivatives.
In~\cite{kim2020simultaneous, kim2021accurate} a simultaneous feedrate optimization and error pre-compensation method is proposed for a system described by a linear model. This approach is analogous to the first stage optimization described in \ref{sec:stage-one}. 

Therefore, to the best of our knowledge, there is a gap in the existing literature on pre-compensation methods for improving PMS performance/productivity that are data-driven, capable of generalizing to novel geometries, and applicable to a broad variety of machines that can only be accuratelly described by a nonlinear model.

In this paper we tackle the challenge of simultaneously improving the performance/productivity of a PMS using only data from the system. Our main contribution is a novel pre-compensation method based on data-driven modelling and trajectory optimisation. 
\if01
\begin{enumerate}
	\item[(i)] A data-driven framework for PMS modelling and identification for control, including design of experiments for collecting training data, and machine learning architecture in a feed-forward form.
	\item[(ii)] A novel contour control pre-compensation method featuring time-optimal trajectory design and motion constraints.
	\item[(iii)] Experimentally validated improvement over the existing controller across the precision/productivity trade-off curve.
\end{enumerate}\fi
Our approach does not require modification of the design or control software of the system (i) and is fully data-driven (ii). Moreover, it enables adaptation to changing conditions or system degradation through retraining the underlying data-driven model (iii). Finally, the approach is independent of the target geometry and significantly shifts the precision/productivity trade-off curve\footnote{Relative to a method without pre-compensation, our proposed method is 30-60\% more accurate for new geometries for same trajectory time budget or alternatively, trajectories can be traced in 25-50\% of the time for the same contouring accuracy.} (iv).

We build upon our previous work \cite{balula2020reference} based on linear system identification. Here we improve and extend this approach by incorporating a nonlinear ANN-based model of system performance, which offers two orders of magnitude higher precision, a refined optimization approach, and  experimental validation of the simulation results.

This paper is organized as follows: 
in Section \ref{sec:ProblemFormulation} we precisely define the problem tackled, while 
in Section \ref{sec:model} we describe the modelling structure, the model fitting strategy, and the accuracy achieved.
In Section \ref{sec:approach} we detail the input trajectory design strategy proposed, specifying the optimization problems.
In Section \ref{sec:results} we show experimental validation for the method with a set of test cases.


\section{Problem Formulation} \label{sec:ProblemFormulation}
\label{sub:TrackingProblem}

Consider a tooltip, e.g., the laser in a laser cutting machine, which traverses a two-dimensional (2D) workspace $\mathcal{W}\subset \reals^2$.
The position of the tooltip as a function of time is denoted by
$	\OutputOne(t) = [x(t)~~y(t)]^\Transpose.$
The ultimate goal in precision motion planning is to have the tooltip trace a spatial target geometry
$	\Objective : [0,S] \to \mc{W} \,,$
where $S$ is the path length. The target geometry $\Objective$ is assumed to be a continuous function parameterized by the path progress variable $s$.
Typical design specifications require the target geometry to be traced with a precision of tens of micrometers; for example, in industrial laser cutting systems a typical precision specification is $20~\mathrm{\mu m}$

\if01
\begin{figure}[t]
	\begin{center}
		\includegraphics[width=\columnwidth]{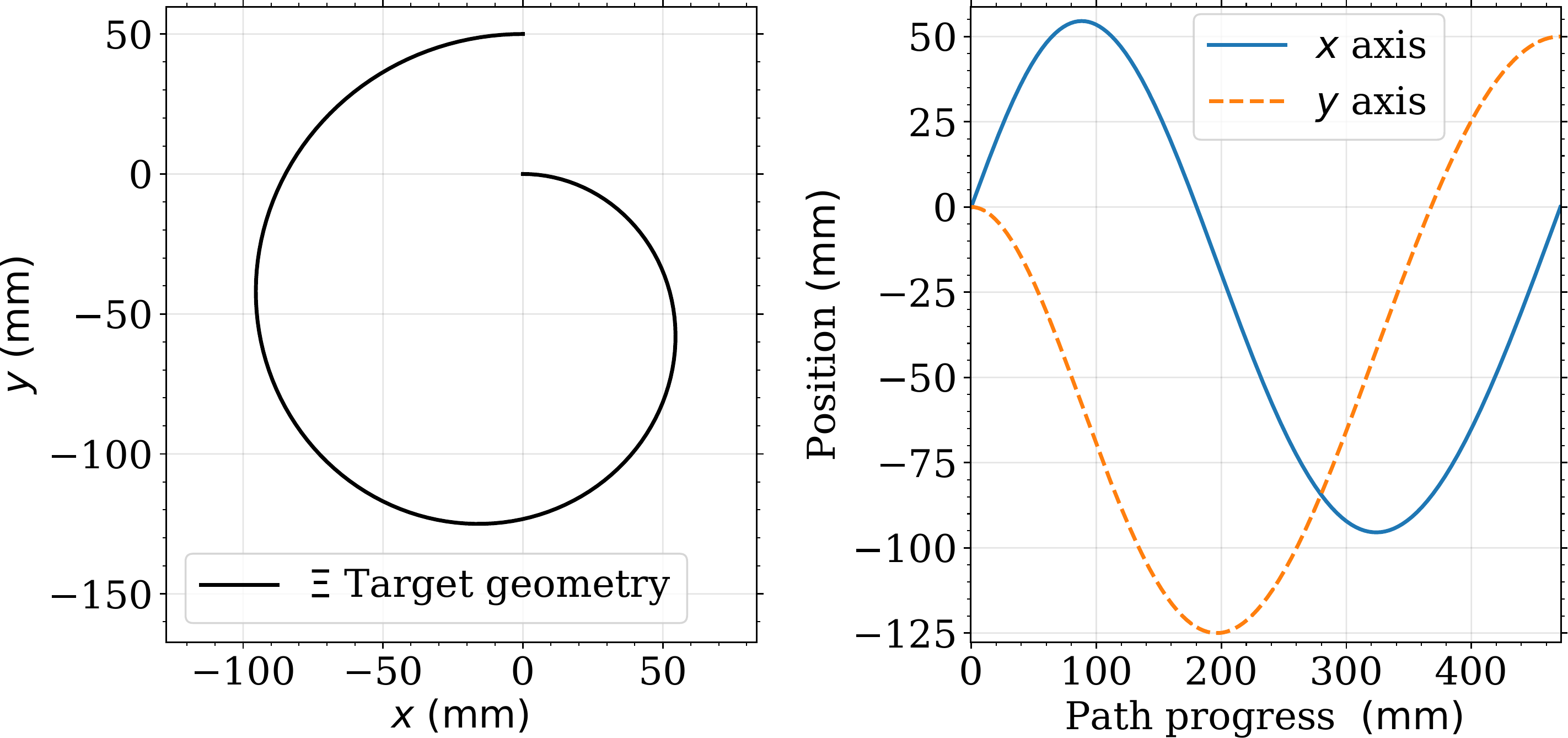} 
		\caption{An example target geometry, featuring a segment of an Archimedean spiral, described in polar coordinates by $r =  b \cdot \theta$ with $b = \frac{50}{2\pi} ~\mathrm{mm\,rad^{-1}}$ and $\theta \in [0, 4\pi]$. The two plots show alternative representations of the same shape. In the left panel it is shown in a $(x(s),y(s))$ plot, parametric in the path progress $s$, while in the right panel each axis is a function of $s$.}
		\label{fig:objective-example}
	\end{center}
\end{figure}
\fi

Most industrial precision motion systems come with a built-in low-level controller designed by the manufacturer to track an input trajectory $\ReferenceOne : [0,T] \to \mc{W}$, where $T > 0$ is the time necessary for the tooltip to complete tracing the target geometry.
When $\ReferenceOne$ is given as an input to the low-level controller, its actions and the dynamics of the machine give rise to an output trajectory $\OutputOne:[0,T] \to \mc{W}$; note that for simplicity we assume that both the input and the output trajectory lie in the workspace $\mc{W}$.
We denote the mapping between the commanded input trajectory $\ReferenceOne$ and the resulting output trajectory $\OutputOne$ by
\begin{equation}
\OutputOne = F(\ReferenceOne) + e,
\end{equation}
where the function $F: \mc{W}^{[0,T]} \to \mc{W}^{[0,T]}$ encapsulates the closed-loop dynamics of the machine. 
The noise $e$ is zero-mean and quantifies the repeatibility of the system, a typical standard deviation is $\approx 2\,\mathrm{\mu m}$.
Ideally, the low-level controller is well designed so that $\OutputOne \approx \ReferenceOne$ and $F$ is approximately the identity function.

We are interested in the following inverse problem: Given the target geometry $\Objective$, how do we generate an input trajectory $\ReferenceOne$ for the machine, so that the resulting output $\OutputOne = F(\ReferenceOne)$ matches $\Xi$ as closely as possible. In addition to the usual difficulties associated with inverse problems, for precision motion systems the function $F$ is often poorly known, partly due to lack of information about the low-level controller.

In standard practice, input trajectories are generated by
traversing the path at a constant speed $v > 0$, i.e. setting $\ReferenceOne(t) = \Xi(vt)$.  However, due to sensor noise, actuator limits, the inertia of the machine, and other factors we observe experimentally a significant error between $\OutputOne$ and $\ReferenceOne$, especially when $v$ is high.
This induces a natural trade-off between speed and precision that is characteristic of precision motion systems. Our aim is to improve this trade-off by learning the function $F$ through preliminary experiments, then inverting it using optimization. \\

\begin{figure}[t]
	\begin{center}
		\includegraphics[width=\columnwidth]{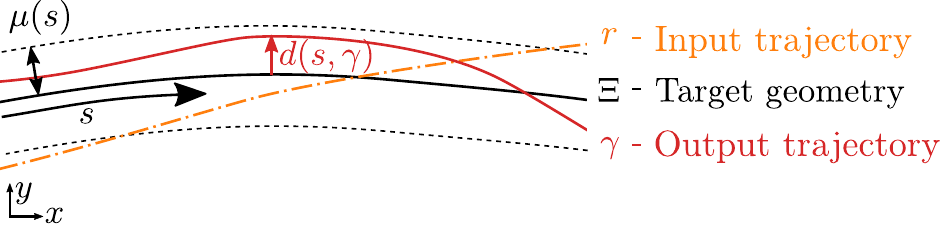} 
		\caption{A low-level controller tracks the input trajectory $\ReferenceOne$, producing the output trajectory $\OutputOne$. The deviation $d(s,\OutputOne)$ measures the distance from $\OutputOne$ to the target geometry $\Xi$ as a function of the path progress $s$. We desire that the deviation is smaller than the tolerance $\mu(s)$ at all points.}
		\label{fig:sketch}
	\end{center}
\end{figure}

Formally, let $d:[0,S] \times \mc{W}^{[0,T]} \rightarrow \reals$ be the deviation function, defined as the projection of the output into the target geometry
$	d(s,\OutputOne) = \min_{t\in [0,T]} \| \Objective(s) - \OutputOne(t) \|_2.$
Our goal is then to solve the optimization problem
\begin{mini}[c1] {\ReferenceOne}{T(\OutputOne)+ \lambda \int_0^S d(s,\OutputOne) ds}{}{}
\addConstraint{\OutputOne =~}{F(\ReferenceOne)}{}{}
	\addConstraint{d(s,\OutputOne) \leq ~}{\Bound(s)}{,~\forall s \in [0,S]} \, , \label{eq:ideal_opt}
\end{mini}
where $T(\OutputOne)$ is the duration of the output trajectory, $\lambda > 0$ governs the trade-off between speed and accuracy, and $\Bound(s): [0,S] \rightarrow \mathbb{R}$ is a tolerance bound around the target geometry; for the typical industrial laser cutting systems mentioned above, one would use $\mu(s) = 20~\mathrm{\mu m}, ~ \forall s\in [0,S]$. The different quantities are illustrated in Figure~\ref{fig:sketch}.

The optimization problem \eqref{eq:ideal_opt} is stated in a trajectory space and is not computationally tractable as written. In the following sections we present and experimentally validate a practical methodology for approximately solving \eqref{eq:ideal_opt} using only data gathered from the process. Because the methodology is data-driven and noninvasive (in the sense that it does not require any knowledge of the underlying controllers but only the ability to run experiments on the machine) it is applicable to a wide range of practical industrial applications.


\subsection{Experimental Setup} \label{subsec:setup}
\begin{figure}[t]
	\begin{center}
		\begin{subfigure}{\columnwidth}
		\includegraphics[width=\textwidth]{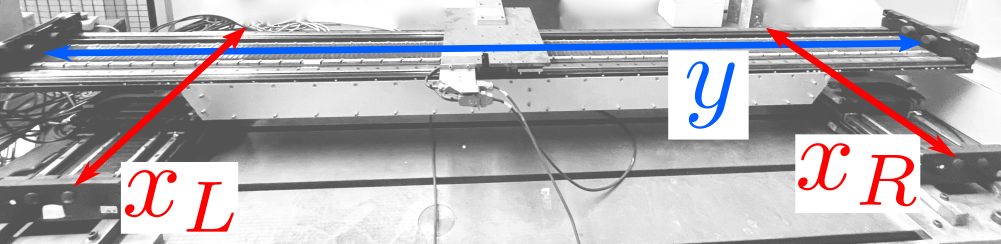} 
			\caption{Experimental apparatus}
					\label{fig:andromeda}
		\end{subfigure}\hfill

		\begin{subfigure}{.49\columnwidth}
			\centering
			\includegraphics[width=.9\textwidth]{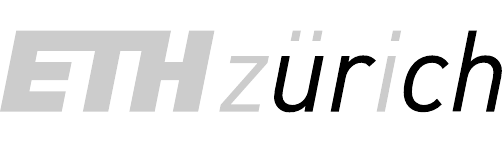}
			\caption{\Letters~test case}
			\label{fig:letters}
		\end{subfigure}\hfill
		\begin{subfigure}{.49\columnwidth}
			\centering
			\includegraphics[width=.9\textwidth]{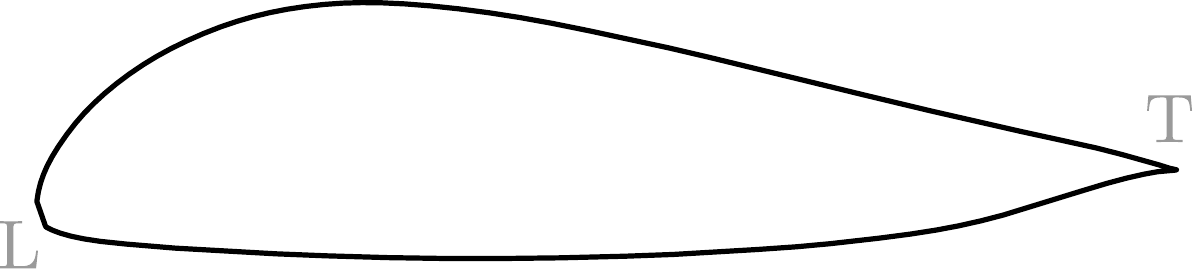}
			\caption{\Airfoil~test case}
			\label{fig:airfoil}
		\end{subfigure}\hfill
		\caption{The experimental setup shown in panel \ref{fig:andromeda} features two degrees of freedom, $x$ and $y$, driven by linear actuators. The end effector is mounted on the $y$ axis, which in turn is mounted on top of the $x$ axis. In panel \ref{fig:letters} the contour of the letters in black is the \Letters~test case, and the outline in panel \ref{fig:airfoil} the \Airfoil~test case, labeled with the leading (L) and trailing (T) edges locations.}
	\end{center}
\end{figure}
While our proposed methodology is applicable for general PMS, we use a specific experimental setup as a running example throughout the paper and to experimentally validate our approach. The experimental setup is shown in Figure~\ref{fig:andromeda}, and consists of an \texttt{ETEL DXL-LM325} two stage motion machine. The machine is instrumented with quadrature encoders that measure the position of both axes and is actuated with \texttt{ETEL LMS15-100} linear motors.

The system is equipped with a feedback controller that adjusts the motor voltages to track a supplied input trajectory $\ReferenceOne$. The controller is implemented on a \texttt{Speedgoat} rapid prototyping unit using \texttt{MATLAB/Simulink} software. For the purposes of this paper we consider this loop as a black-box and consider only the desired input trajectory $\ReferenceOne$ as an input.
The machine enforces limits on the acceleration, velocity, and position of the input trajectory to prevent accidental damage. The main characteristics of the machine are collected in Table~\ref{tab:machine-limits}.

\begingroup
\renewcommand{\arraystretch}{1.3}
\begin{table}
	\begin{center}
		\begin{tabular}{r c l}
			\toprule
			Property	&Value 	&Unit\\
			\midrule
			Maximum acceleration 			&$40$&$\mathrm{m\,s^{-2}}$\\  
			Maximum velocity 				&$1.5$&$\mathrm{m\,s^{-1}}$\\ 
			Working area					&$[-0.19, 0.19]^2$&$\mathrm{m^2}$\\
			Low-level control loop frequency 	&$10$&$\mathrm{k Hz}$\\
			Repeatability (standard deviation) &$2$&$\mathrm{\mu m}$ \\
			\bottomrule
		\end{tabular}
		\caption{\label{tab:machine-limits}Software-limited values and operational specifications of the experimental setup.}
	\end{center}
\end{table}

\endgroup

\subsection{Test cases} \label{testcases}
We selected two test cases, a series of letters and an airfoil, as target geometries to validate our proposed methodology. These geometries contain a rich set of features that are representative of industrial applications. To avoid ``cherry picking'', and to demonstrate generalization to new geometries, neither test case was used to generate training data or during the development of the method.

The \textbf{letters} test case, as can be seen in Figure \ref{fig:letters}, consists of four letters (``u'',``r'',``c'', and ``h'') from the ETH Z\"{u}rich logo. These letters were selected due to the rich set of relevant and challenging features they contain, including sharp corners, straight segments, and curves with various curvatures. The limiting contour of each letter is used as the target geometry.

The \textbf{airfoil} test case, shown in Figure \ref{fig:airfoil}, is based on a high-lift Eppler e344 airfoil\cite{eppler1990airfoil}. Geometries of this kind are typically laser cut out of e.g., balsa wood, and are used in the primary structural sub-assembly of an aircraft. The main challenging features of the airfoil are a cut-out at the leading edge (L) used to attach a structural spar, and the trailing edge (T) whose geometry has a strong effect on the aerodynamics of the aircraft.

\subsection{Architecture overview}\label{sec:architecture}
\begin{figure}[t]
	\begin{center}
		\includegraphics[width=\columnwidth]{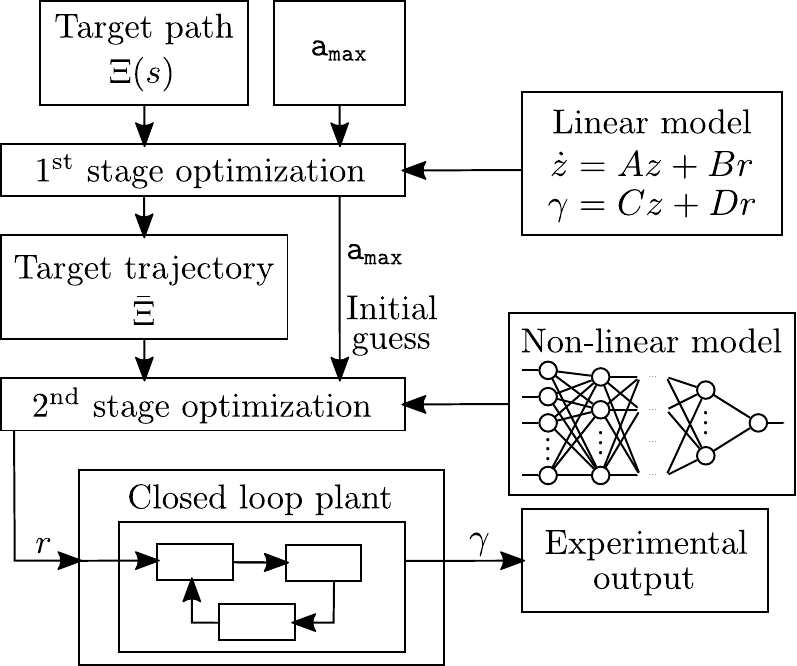} 
		\caption{
		In the first stage optimization \eqref{eq:stage1} a minimum time trajectory for tracking the target path $\Objective(s)$ is computed. 
		The speed and acceleration profile from the first stage is used to sample the target geometry at constant sample rate of $400~\mathrm{Hz}$, creating a surrogate target discrete time trajectory $\bar \Xi$.
		In the second stage optimization \eqref{eq:stage2} nonlinear model is used to find the best input trajectory $\ReferenceTwo$ to track $\bar \Xi$.
		The maximum allowable acceleration $\mathtt{a_{max}}$ governs the accuracy/trajectory time trade-off. 
		The optimized input trajectory $\ReferenceTwo$ is given to the experimental apparatus. 
		}
		\label{fig:scheme}
	\end{center}
\end{figure}

We approach the trajectory design problem in several stages. First, we design experiments to generate informative input-output data for the unknown mapping $F$ and use the resulting data to identify two models for the system, a low-fidelity linear one and a high-fidelity one based on an ANN. Throughout, we incorporate machine learning best practices,
such as independent training and validation data sets, normalization, and diverse training data to avoid over fitting and 
to ensure that the resulting models are able to generalize to trajectories outside the training data set, enabling optimization of new geometries. 

Using these models, we propose a two-stage optimization-based approach for offline optimization of references as schematized in Figure~\ref{fig:scheme}. In the first stage, we compute an input trajectory by solving a contouring control problem using the linear model of the system. This first stage solution yields a fast trajectory that respects the problem constraints, by reducing the speed in intricate features of the path, such as sharp corners, and accelerating through long smooth segments. The output of the first-stage is used as the initial condition for the second stage, which employs the high-fidelity neural network model to correct errors from the first stage. The resulting input trajectory is then given to the low-level controller as an input trajectory.

\section{Data driven modelling} \label{sec:model}
As a first approximation, we model the system using a continuous-time linear model which captures the dominant dynamics. The structure of this model allows for efficient computations of a time-optimal solution in the first stage, but lacks the high-precision required by the application. This shortcoming is alleviated by a second high-fidelity neural network model, used to refine the initial solution. 

\subsection{Low-fidelity linear model} \label{sec:linear-model}

The closed-loop system maps trajectories to trajectories and is of the form $\OutputOne = F(\ReferenceOne)$. We approximate this mapping using a linear time invariant (LTI) state space model of the following form
\begin{subequations} \label{eq:linearmodel}
	\begin{align}
		\dot z(t) &= A z(t) + B \ReferenceOne(t) \mathrm {,} \label{eq:linearmodel-a}\\
		\OutputOne(t) &= C z(t) + D \ReferenceOne(t) \mathrm{.} \label{eq:linearmodel-b}
	\end{align}
\end{subequations}
The model \eqref{eq:linearmodel} uses a non-physical hidden state $z \in \reals^4$; the matrices $A$, $B$, $C$, and $D$ are identified from experimental data using the \verb|MATLAB| function \verb|n4sid| from the system identification toolbox. The dimension of the state was chosen empirically, noting that increasing the model order further has diminishing returns.


\subsection{High-fidelity Nonlinear Model} \label{sec:nonlinear-model}

To obtain a higher precision model, 
%
%
the 2D-positioning stage is modeled using two independent causal ANN, one for each axis. Unlike \eqref{sec:linear-model} the nonlinear model operates in discrete time, at a sample frequency of $400\,\mathrm{Hz}$ and it has the following form
\begin{equation}
	\OutputTwo_{i} = f_{nlm}(\ReferenceTwo_{i-h}, \dots, \ReferenceTwo_{i-1}),\label{eq:nlm}
\end{equation}
where $\OutputTwo$ is the output,
$f_{nlm}$ is the nonlinear function defined by the ANN,
$\ReferenceTwo$ is the input trajectory given to the closed loop system,
and $h$ is the length of the input history considered for the prediction.
\if!\ReferenceOne\ReferenceTwo
Note that we have introduced the variables $\ReferenceTwo$ in place of $\ReferenceOne$, and $\OutputTwo$ in place of $\OutputOne$ to make explicit that $\ReferenceOne$ and $\OutputOne$ are parameterized in continuous time, while $\ReferenceTwo$ and $\OutputTwo$ are defined in discrete time.
\fi
\if01
\begin{figure}
	\begin{center}
		\includegraphics[width=\columnwidth]{neuralnetwork/neuralnetwork} 
		\caption{Neural network structure. Each layer is fully connected to the next. The input layer has $200$ neurons, there are $8$ hidden layers, with $\{200, 200, 100, 100, 50, 25, 12, 6\}$ neurons, and the output layer has $1$ neuron. The activation function of the hidden layers is a $\mathrm{LeakyReLU}$ function with slope $0.01$, and a bias term. There are total of $117322$ weights to be adjusted in the training. Two independent networks with identical structure were trained, one for each axis.}
		\label{fig:nn}
	\end{center}
\end{figure}
\fi
The two ANN are structurally identical, but are trained separately, leading to different weights. Each takes as input the most recent $500 \mathrm{m s}$ of the input trajectory for the corresponding axis, subsampled at $400 \mathrm{Hz}$, leading to $200$ inputs. Each network has a single output, namely the predicted position at the subsequent time step. The networks have $8$ fully connected hidden layers of LeakyReLU~\cite{masetti2020analyzing} activation functions with a slope of $0.01$, and a triangular structure. 
The length of the time history was selected to be approximately $25$ times the timescale of the slowest mode of the identified linear model. The sample rate was selected by analyzing how much distance can be covered at the highest expected speeds, and comparing it to the desired range of precisions. The number of hidden layers and their dimensions were selected empirically, to achieve good performance with the lowest possible complexity.
The choice of feed-forward 
ANN was made to ensure that the model prediction accuracy does not degrade over time due to compounding errors. We also note that by training separate ANN for each axis we are implicitly ignoring coupling effects between the two axes; this choice is supported by experimental data that suggests coupling effects are negligible, as will be demonstrated in Section~\ref{sec:num-results}.

\begin{figure}
	\begin{center}
		\includegraphics[width=.8\columnwidth]{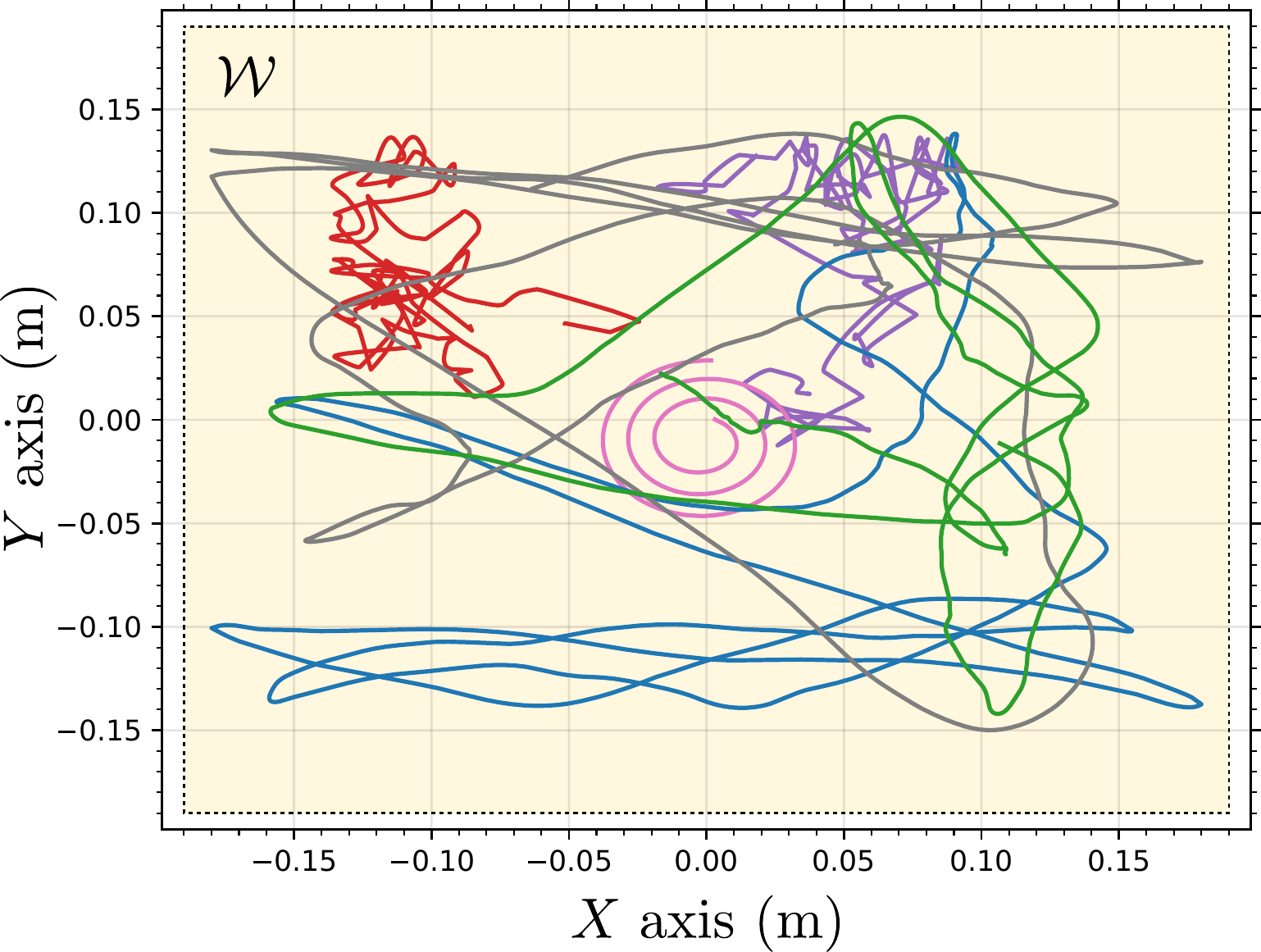} 
		\caption{Sample of the data collected to train and test the nonlinear neural network model. 
		Six trajectories are shown, five randomly generated and one regular shape (a spiral). The trajectories are contained in the workspace $\mathcal{W}$, shown in yellow background. 
		}
		\label{fig:xy-training-data}
	\end{center}
\end{figure}

\subsection{Training data generation}
Choosing input trajectories that cover a wide range of operational regimes is essential for generating informative data to train models that can generalize to previously unseen trajectories.

\subsubsection{Linear Model}\label{paragraph:linear-model-data}
For the identification of the linear model we opted for randomly generated trajectories, yielding $6\times10^{6}$ points in total. By randomly changing the velocity, acceleration, and jerk of the input trajectory we made sure that the trajectories represent features seen in real parts. This included, for example, curves of different radii, traversed at different speeds and appearing in different locations within the workspace, as well as sharp transitions in acceleration and jerk. The trajectories were not designed for a specific part, but for a particular set of constraints on velocity and acceleration.

\subsubsection{Nonlinear Model}\label{paragraph:nonlinear-model-data}
Having sufficient high-quality data is necessary to ensure that the ANN is accurate and generalizes well. 
We started by training a first generation ANN with the random trajectories used to fit the linear model, and a set of varied regular geometries including circles, polygons, and spirals. Each shape was traversed at varied speeds, constant for each trial.
To enrich the data set further, the trained first generation ANN was used in the optimization methodology of Section \ref{sec:approach} applied to the same regular geometries. This resulted in optimized input trajectories and, after applying these to the system, additional experimental data. This data was used to train the ANN further obtaining a second generation ANN.
This training-optimization-retraining cycle can be executed recursively until the model prediction accuracy is acceptable.
\if01
\begin{figure}
	\centering
	\includegraphics[width=\columnwidth]{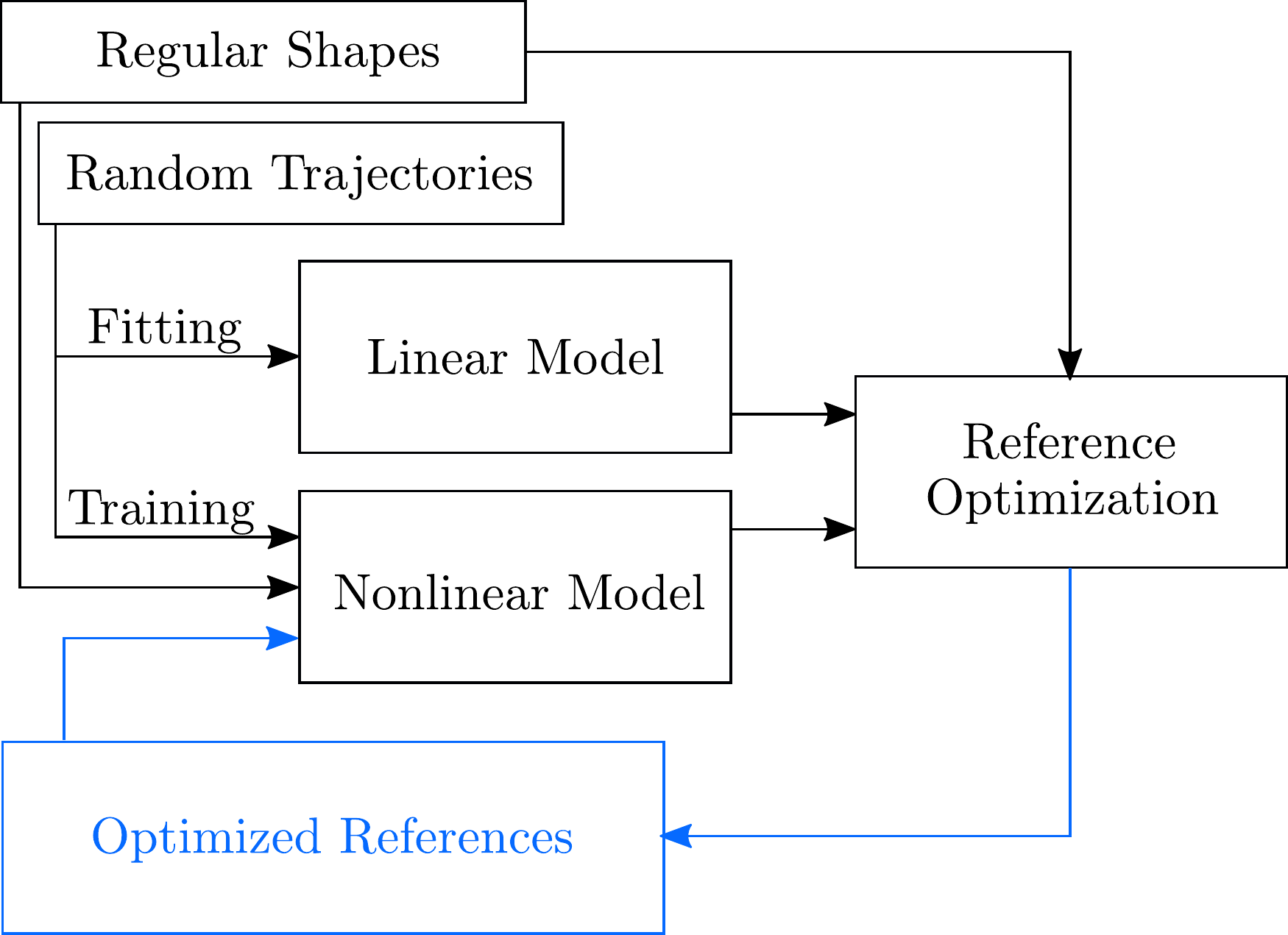}
	\caption{Flow of data used to train the linear and nonlinear models. First experimental data is collected for randomly generated trajectories and regular shapes. This data set is used to fit and train the linear and a first generation nonlinear models. The models are used to compute optimized input trajectories for the set of regular shapes using the method of Section \ref{sec:approach} the additional data is used to extend the training the nonlinear model. This is done because the optimizer has a tendency to exploit regions of the nonlinear model that are not well fitted to reality.}
	\label{fig:data-flow}
\end{figure}
\fi
These data sets lead to a targeted reduction in over fitting in areas that are favored by the optimization. Overall, our strategy is to include sufficient random (exploratory) and structured (exploitative) data to ensure the ANN is both accurate and generalizes well in the areas of interest.

A representative training data sample is shown in Figure \ref{fig:xy-training-data}
In section \ref{sec:num-results} we present the prediction quality of a third generation ANN to generate the results of Section \ref{sec:results}.
The data was divided $80\%/20\%$ for training and testing purposes.
This results in a ratio of $\approx34$ data points per parameter of the ANN. About $60\%$ of the data comes from the randomly generated trajectories, and the remaining from regular shapes both before and after optimization. The experimental data is divided in segments of $500~\mathrm{ms}$, subsampled to achieve an effective sample rate of $400\mathrm{Hz}$, and scaled.
For the loss function we use the mean square of the prediction error.
We use batches of $16\mathrm{k}$ segments which take maximum advantage of the GPU computation power. {\it Adam}~\cite{kingma2014adam} was used for the optimizer, with decaying learning rate. Modeling and training were done in {\it PyTorch}~\cite{NEURIPS2019_9015} with {\it NVIDIA CUDA} support. 

\subsection{Model Validation} \label{sec:num-results}
\newcommand{\umph}{}
\newcommand{\umphb}{\mathrm{\mu m}}
\begingroup
\renewcommand{\arraystretch}{1.3}
\begin{table}
	\begin{center}
		\begin{tabular}{@{}*{8}{r}@{}}
			\toprule
			&&\multicolumn{2}{c}{Training / } 	&\multicolumn{2}{c}{$\mathtt{a_{max}}$} &\multicolumn{2}{c}{$\mathtt{a_{max}}$} \\
			&&\multicolumn{2}{c}{Validation} 	&\multicolumn{2}{c}{$1.0~ \mathrm{m\,s^{-2}}$} &\multicolumn{2}{c}{$3.0~ \mathrm{m\,s^{-2}}$} \\
			\cmidrule(lr){3-4} \cmidrule(lr){5-6}  \cmidrule(lr){7-8} 			
			Model&Axis	&$\mu\umph$				&$\sigma\umph$  &$\mu\umph$			&$\sigma\umph$			&$\mu\umph$				&$\sigma\umph$\\
			--&--&$\umphb$	&$\umphb$	&$\umphb$	&$\umphb$	&$\umphb$	&$\umphb$	\\
			\midrule 
			Linear		&x		&3.0&48.0		&4.7&30.1	&5.4 &57.2 \\
						&y		&-13.2&156.3	&0.0&108.5	&0.5&229.4 \\
			\midrule
			Non-linear	&x		&0.0&3.2		&-0.3&8.2	&1.0&11.1	\\
						&y		&0.0&6.0		&-2.0&13.4	&-2.3&18.7	\\
			\bottomrule
		\end{tabular}
		\caption{\label{tab:linear-model-acc} Prediction error mean $\mu$ and standard deviation $\sigma$ of the linear and non-linear models. Both models are evaluated in the \Letters~test case after optimization, with different maximum acceleration values $\mathtt{a_{max}}$. The linear model is also evaluated on its training data set, and the non-linear model on its validation data set.}
	\end{center}
\end{table}
\endgroup

\if01
\begin{figure}
\centering
\begin{subfigure}{\columnwidth}
	\includegraphics[width = .49\textwidth]{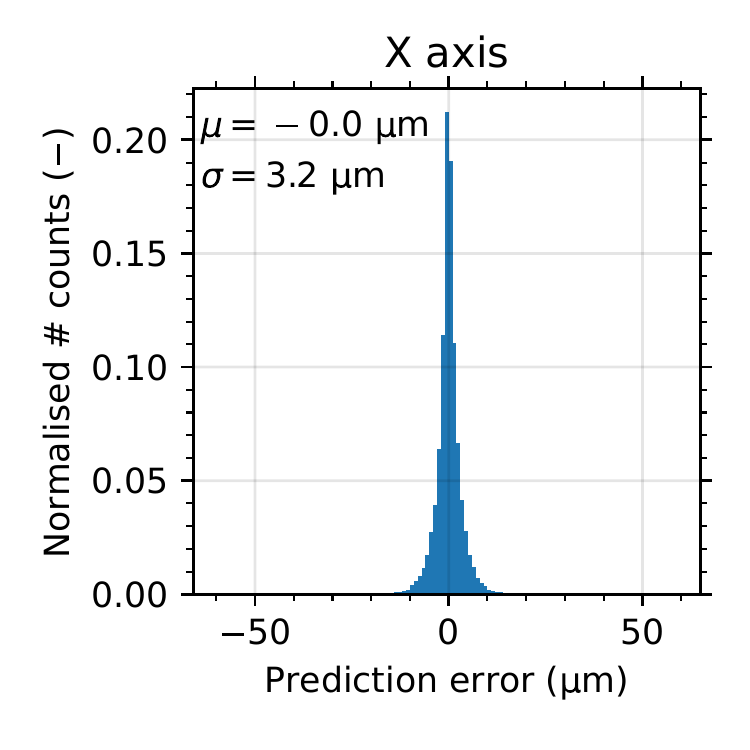}
	\includegraphics[width = .49\textwidth]{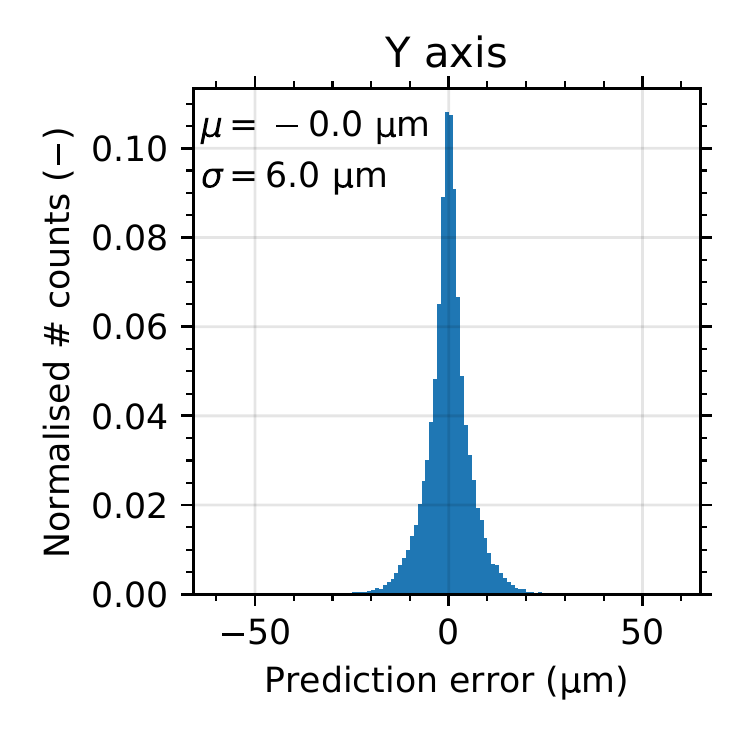}
  \caption{Validation data\label{fig:label1}}
\end{subfigure}\hfill
\begin{subfigure}{1\columnwidth}
	\includegraphics[width = .49\textwidth]{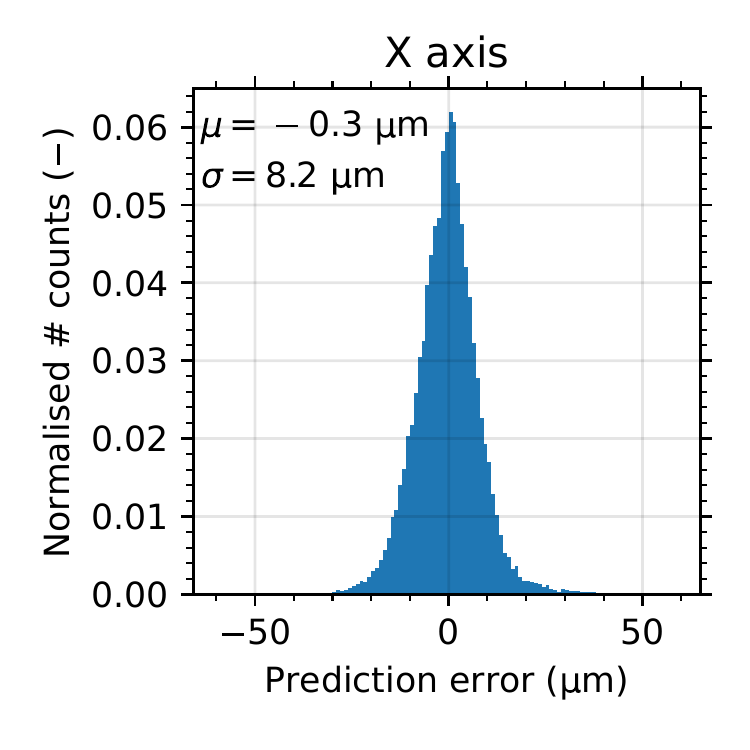}
	\includegraphics[width = .49\textwidth]{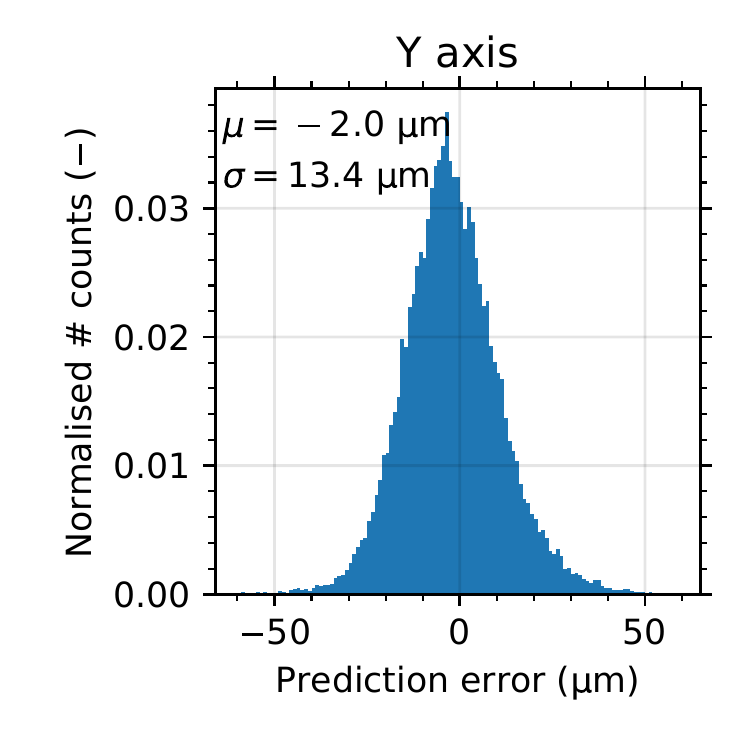}
	\caption{\Letters~test case with $\mathtt{a_{max}} \leq 1~\mathrm{m\,s^{-2}}$}
  \label{fig:label2}
\end{subfigure}\hfill
\begin{subfigure}{\columnwidth}
	\includegraphics[width = .49\textwidth]{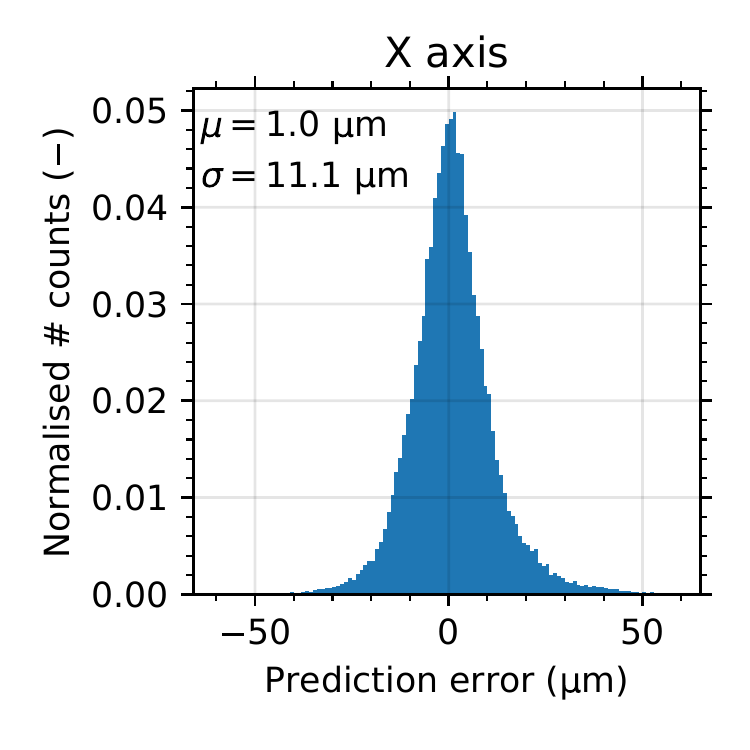}
	\includegraphics[width = .49\textwidth]{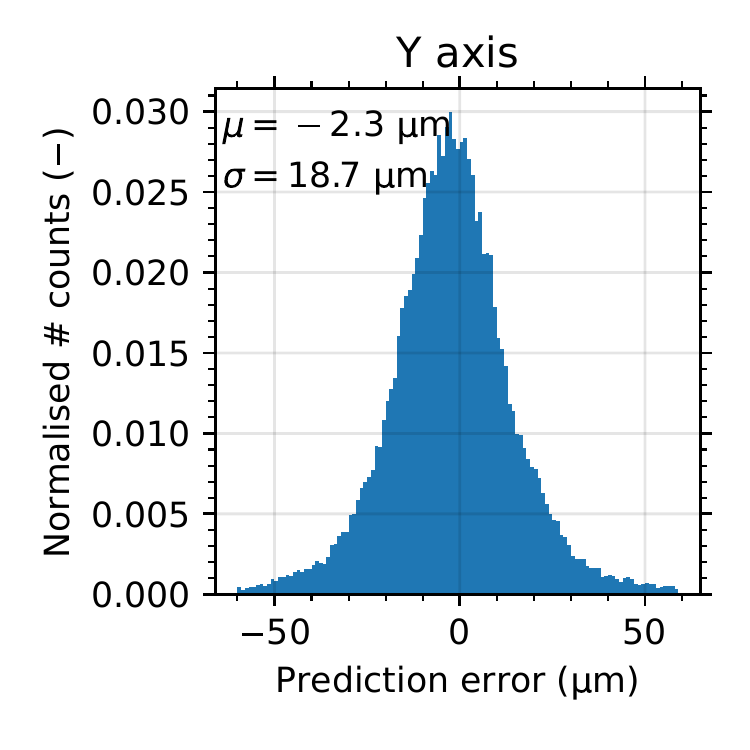}
	\caption{\Letters~test case with $\mathtt{a_{max}} \leq 3~\mathrm{m\,s^{-2}}$}
  \label{fig:label3}
\end{subfigure}
	\caption{Histogram of the prediction error of the non-linear model. In panel (a) 
	the model is evaluated in the validation data set, as exemplified in Figure~\ref{fig:xy-training-data}.
	In panels (b) and (c) 
	the model is evaluated in the \Letters~test case, with the input trajectory optimized with $\mathtt{a_{max}}$ up to $\{1, 3\}~\mathrm{m\,s^{-2}}$.}
\label{fig:model-hist}
\end{figure}
\fi

\begin{figure}
	\begin{center}
		\includegraphics[width=\columnwidth]{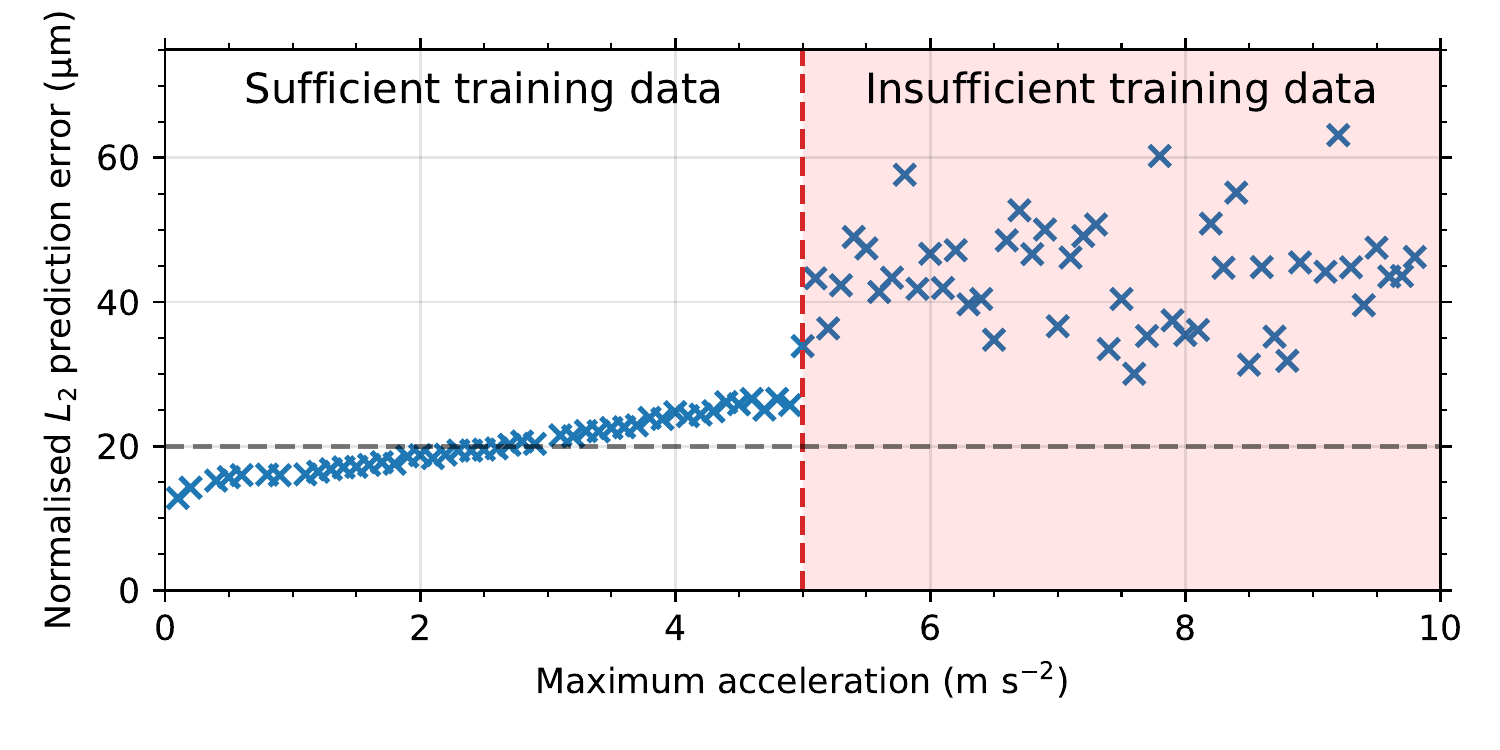} 
		\caption{Prediction error of the nonlinear model as a function of the maximum acceleration $\mathtt{a_{max}}$ of the input trajectory. The error is evaluated for the letters and airfoil test cases. Above $5~\mathrm{m\,s^{-2}}$ the training data do not contain enough information to train the model properly.
		}
		\label{fig:opterror-vs-acc}
	\end{center}
\end{figure}

The quality of the model predictions is evaluated over different data sets and acceleration limits. The results are summarized in Table \ref{tab:linear-model-acc}.

The linear model training data set includes trajectories with a range of accelerations up to $5~\mathrm{m\,s^{-2}}$. However, due to its simple structure, the linear model cannot overfit the data as it does not have the representative power to provide accurate predictions everywhere in its training set. We observed that it performs better when the acceleration is restricted to a lower value, even for trajectories not seen during the fit, such as the letters test case. The linear model accuracy is insufficient for the demanding requirements of the application.

Regarding the nonlinear model, the prediction accuracy is best for data with the same characteristics as the one used for the training. 
When presented with data coming from shapes which have not been used for the training of the model, 
the prediction accuracy degrades. However, the accuracy is still adequate for the application, as the standard deviations remain below $20~\mathrm{\mu m}$ for each axis for accelerations below $3~\mathrm{m\,s^{-2}}$. 

Figure \ref{fig:opterror-vs-acc} shows how the standard deviation of the prediction evolves with the maximum acceleration $\mathtt{a_{max}}$ that was set as a constraint in the optimization of Section \ref{sec:approach}. For acceleration values where there is sufficient training data, the model prediction error increases gradually with the acceleration. If we try to use the model outside of the acceleration values it was trained for, the predictions rapidly deteriorate.

\begin{figure}[t]
	\begin{center}
		\includegraphics[width=\columnwidth]{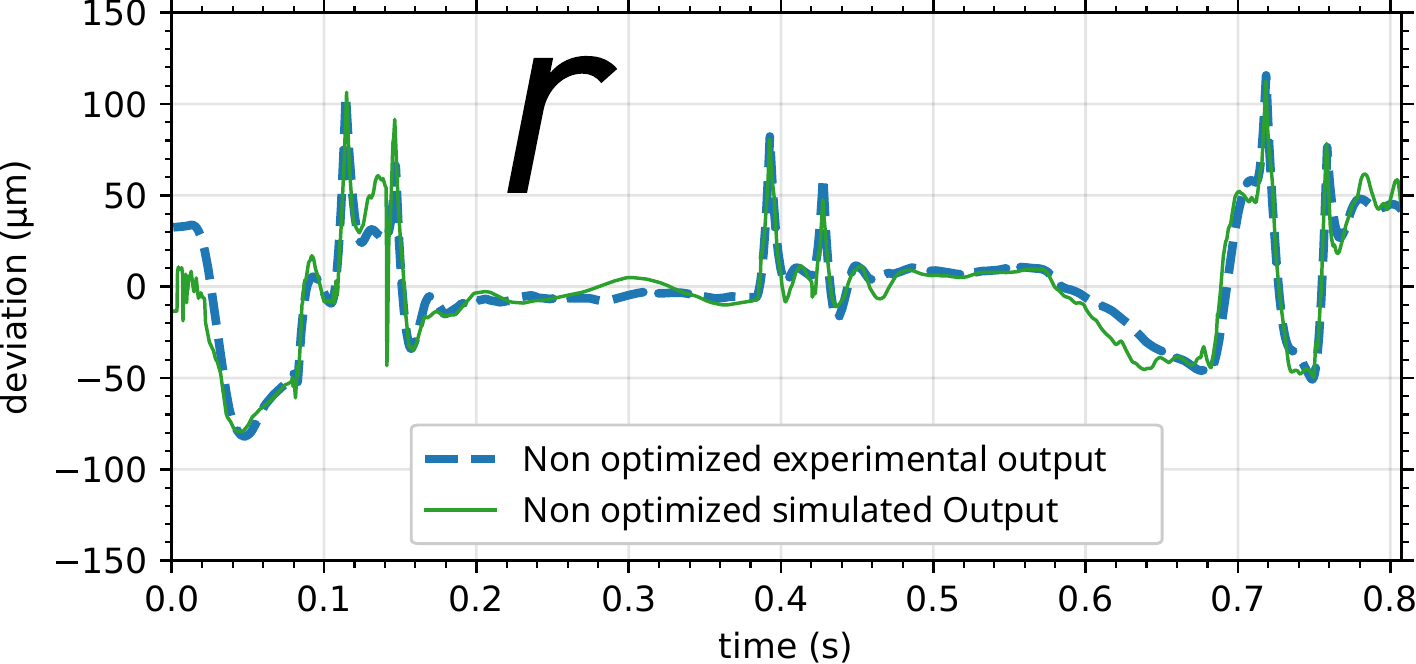} 
		\caption{Deviation from target geometry of simulated and experimental output. The target geometry of the letter ``r'' of the ETH Z\"{u}rich Logo, traveled at a constant path speed of $40~\mathrm{mm\,s^{-1}}$ is used as the reference. The simulated output corresponds to the prediction of the nonlinear model, which is plotted with the experimental data collected with the same reference. Both simulated and experimental results are compared against the target geometry to compute the deviation. The model predictions follow the experimental data closely. This plot also shows the performance of the low-level controller, with deviations exceeding $100~\mathrm{\mu m}$.}
		\label{fig:r-time}
	\end{center}
\end{figure}

\if01
A representative predicted trajectory from the nonlinear model is shown in Figure~\ref{fig:r-time}, in which the letter ''r`` from the letters test case was simulated by using as reference the target geometry at a constant path speed.

\fi

\section{Trajectory Optimization} \label{sec:approach}
With the the data-driven models in hand, we now discuss the two-stage optimisation architecture used to compute input trajectories. The decision to divide the optimization problem in two stages is motivated by the need to include the duration $T$ of the trajectory in a computationally tractable manner. This is accomplished by using a variable time-discretization
in the first stage, enabling inclusion of $T$ as a decision variable. The second stage then fixes the duration $T$ and refines the accuracy of the first-stage solution using the high-fidelity neural network model. 

Our experiments suggest that treating the time discretization directly with the nonlinear model is intractable. A key difficulty is that real machines typically generate noisy output measurements in a sampled, time series format. These cannot be readily used to train the high-fidelity continuous time model needed for a computationally tractable variable-time discretization, due the challenges associated with numerical differentiation of noisy data. Trying to solve the inverse problem in one shot using the resulting model leads to low quality solutions; often the solver fails to even return a feasible solution in a reasonable amount of time.

On the other hand, skipping the second stage and inverting using only the linear model is computationally tractable, but typically leads to low quality solutions; this observation is supported by the model prediction errors shown in Table \ref{tab:linear-model-acc}. Training a fixed sampling time nonlinear model and seeding with the solution of the first stage optimization substantially reduces the overall computation time and improves the quality of the solutions.

A key advantage of our methodology is that, unlike iterative learning control, once the models are trained and we are given a target geometry, the input trajectory is computed off-line, without the need for additional experiments. This is especially important for small batch manufacturing where it is crucial to minimize failed parts.
However we can still use the information collected from experimental runs to improve the model. 

\subsection{First-stage Optimization Formulation} \label{sec:stage-one}
The purpose of the first stage is to fix the duration of the trajectory, $T$, by trading off speed and precision. We employ a contouring control approach with a fixed spatial discretization. This involves sampling the target geometry at $N$ equally spaced points in space $\{\Objective_k \doteq \Objective(s_k): k\in\{0,1,\dots,N-1\}\}$, where
$s_k = k \frac{S}{N-1} = k \Delta s.$

We use the identified linear model \eqref{eq:linearmodel} to approximate the process dynamics. As our approach involves a fixed spatial discretization $\Delta s$, the time-discretization must be variable.
Starting with $t_0=0$, we denote by $t_k$ as the time at which we would like the trajectory to reach the point $\Objective_k$ and consider the non-uniform time steps $\DeltatOne_k = t_{k+1} - t_k, k=\{0, \dots, N-1\}$ as decision variables; note that the time $T=t_N=\sum_{k=0}^{N-1} \DeltatOne_k$ at which the trajectory is completed is implicitly also a decision variable.
We then obtain a discrete time model by applying the $4$th order Runge-Kutta (RK4)~\cite{aboanber2008generalized} time-marching method to \eqref{eq:linearmodel}
\begin{subequations}
	\label{eq:fnlm}
\begin{align}
	z_{k+1} &= f_{\mathrm{lm}}(z_k, \ReferenceOne_k, \DeltatOne_k)\\
\OutputOne_{k} &= C z_k + D \ReferenceOne_k,
\end{align}
\end{subequations}
where $f_{\mathrm{lm}}$ is the RK4 function, and $\DeltatOne$ will be treated as decision variables.

\begin{figure}
	\centering
		\includegraphics[width=.55\columnwidth]{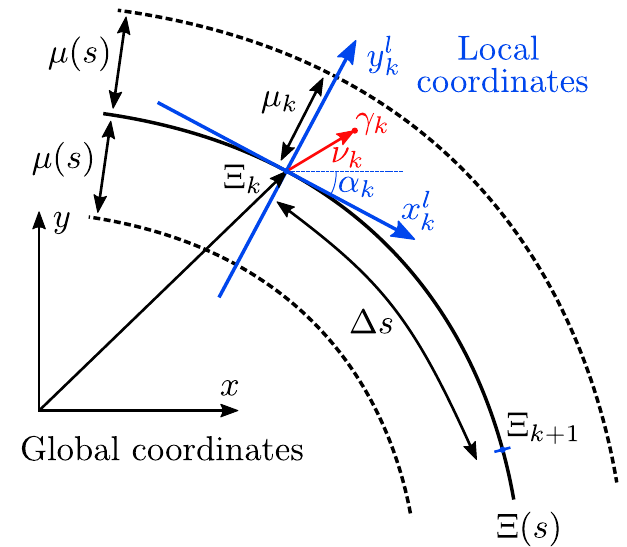} 
		\caption{For every point $\Objective_k$ of the target geometry $\Objective(s)$ a local coordinate frame is placed. $x^l$ is tangent to the target geometry at $\Objective_k$ and pointing in the direction of increasing $s$ whereas $y^l$ is orthogonal to $x^l$ pointing to the port side. $\alpha_k$ is the angle between the global $x$ axis and $x^l$. The maximum allowed deviation $\Bound_k$ is computed from $\Bound(s)$. $\Delta s$ is the path distance between points $\Objective_k$ and $\Objective_{k+1}$.}
		\label{fig:local-var}
\end{figure}

To evaluate the contouring error $d_k$
we start with the distance $\nu_k = \OutputOne_k - \Objective_k$ and introduce a local coordinate frame as shown in Figure \ref{fig:local-var}. The linear transformation
\begin{equation}\label{eq:Tk}
	T_k = \begin{bmatrix}
		\cos{\alpha_k}		&\sin{\alpha_k}	&\Xi_{k,x} \\
		-\sin{\alpha_k}	&\cos{\alpha_k}	&\Xi_{k,y} \\
		0					&0					&1
	\end{bmatrix},
\end{equation}
where $\alpha_k$ is the rotation angle of the local frame, 
encodes a translation and rotation between the local coordinate frame $k$ and the global coordinate frame. With 
\begin{equation} \label{eq:local-global-transformation}
	\begin{bmatrix} \OutputOne_k \\ 1 \end{bmatrix} = T_k \begin{bmatrix} \LocalVar_k \\ 1 \end{bmatrix},
\end{equation}
	the output can be transformed from local coordinates $\LocalVar_k$ to global coordinates $\OutputOne_k$.
Since time discretization is variable, we can always ensure that the $x^l$ component of $\nu_k$ is zero by adding a constraint to the optimization problem. The contouring error is then simply the $y^l$ component of $\nu_k$.
The tolerance bound condition in \eqref{eq:ideal_opt} can then also be computed for the corresponding discretization point $\Bound_k = \Bound(s_k)$.

The aim of the optimization problem is to minimize the total time needed to complete the target geometry subject to the tolerance bounds, velocity and acceleration limits.
To this basic formulation, we introduce an additional term for the input to promote smoothness; this is used to suppress high frequency content on the input trajectory, which is filtered out by the linear model.
This results in the following optimization problem
\begin{mini!}{\DeltatOne, \ReferenceOne, \OutputOne, z, \LocalVar}{\label{eq:stage1-a} \sum_{k=1}^{N-1} \DeltatOne_k + \frac{1}{N-2} \sum_{k=2}^{N-1} \|\partial^2 \ReferenceOne_k\|_2^2}{\label{eq:stage1}}{}
	\addConstraint{\label{eq:stage1-b} z_{k+1}=~}{f_{\mathrm{lm}}(z_k, \ReferenceOne_k, \DeltatOne_k),}{~k = 1, \dots, N-1}
	\addConstraint{\label{eq:stage1-c} \OutputOne_k=~}{C z_k + D \ReferenceOne _k,}{~k = 1, \dots, N}
	\addConstraint{\label{eq:stage1-d} \DeltatOne_k \geq ~}{0,}{~k = 1, \dots, N}
	\addConstraint{\label{eq:stage1-e} \OutputOne_k \in ~}{ \mathcal{W},}{~k = 1, \dots, N}
	\addConstraint{\label{eq:stage1-f} \begin{bmatrix} \OutputOne_k \\ 1 \end{bmatrix}=~}{T_k \begin{bmatrix} \LocalVar_k \\ 1 \end{bmatrix},}{~k = 1, \dots, N}
	\addConstraint{\label{eq:stage1-g} {\LocalVar_k}_1 = ~}{0,}{~k = 1, \dots, N}
	\addConstraint{\label{eq:stage1-h} | {\LocalVar_k}_2 | \leq ~}{	\Bound_k,}{~k = 1, \dots, N}
	\addConstraint{\label{eq:stage1-i} \| \partial \OutputOne_k \|_\infty \leq ~}{ \mathtt{v_{max}},}{~k = 1, \dots, N-1}
	\addConstraint{\label{eq:stage1-j} \| \partial^2	\OutputOne_k \|_\infty \leq ~}{ \mathtt{a_{max}},}{~k = 2, \dots, N-1}
	\mathrm{,}
\end{mini!}
where $N$ is the number of discretization points, selected based on the total path length and tolerance bound,
$\DeltatOne = \{\DeltatOne_k\}_{k=1}^{N-1}\subseteq \reals$,
$
	\partial \OutputOne_k = \frac{\OutputOne_{k+1} - \OutputOne_k}{\Delta \tau_k} \text{ and }
	\partial^2 \OutputOne_k = \frac{\OutputOne_{k+1} - 2 \OutputOne_{k} + \OutputOne_{k-1}}{\Delta \tau_k \Delta \tau_{k+1}}$
are discrete derivative operators, 
$\LocalVar = \{\LocalVar_k\}_{k=1}^{N}\subseteq \reals^2$ is the output trajectory in local coordinates,
$\OutputOne = \{\OutputOne_k\}_{k=1}^{N}\subseteq \reals^2$ is the output trajectory in global coordinates,
$\ReferenceOne= \{\ReferenceOne_k\}_{k=1}^{N-1}\subseteq \reals^2$ is the input trajectory,
$f_{\mathrm{lm}}$ is the RK4 approximation of the linear model \eqref{eq:fnlm}, with identified matrices $C$ and $D$, 
$T_k$ is the transformation matrix between global and local coordinates in \eqref{eq:Tk},
$\Bound_k$ is the discretized maximum allowed deviation from the target geometry,
$\mathcal{W}$ are the limits of the working space of the device,
$\mathtt{v_{max}}$ is the maximum allowable speed, and
$\mathtt{a_{max}}$ is the acceleration limit.

The cost \eqref{eq:stage1-a} contains two terms. The first penalizes the total time to perform the trajectory while the second penalizes the acceleration of the input, promoting fast trajectories with a smooth input,
\eqref{eq:stage1-b} and \eqref{eq:stage1-c} imposes the RK4 discretization of the linear dynamics \eqref{eq:fnlm},
\eqref{eq:stage1-d} requires the variable time step to be non-negative,
\eqref{eq:stage1-e} imposes that the output stays within the working space of the device,
\eqref{eq:stage1-f} the transformation between local and global coordinates in \eqref{eq:local-global-transformation},
\eqref{eq:stage1-g} requires that the $x_k^l$ component of $\nu_k$ is zero, in which case the $y_k^l$ component is the deviation,
\eqref{eq:stage1-h} bounds the $y_k^l$ component of $\nu_k$ to be less than the maximum allowed deviation,
\eqref{eq:stage1-i} and \eqref{eq:stage1-j} impose component-wise limits on velocity and acceleration of the output.
This formulation is analogous to~\cite{kim2020simultaneous} where the feedrate and contour error are optimized using a linear model.

\subsection{Second-stage Optimization Formulation} \label{sec:stage-two}
Experiments applying the trajectory $\ReferenceOne$ generated by \eqref{eq:stage1} directly to the machine suggest that this generally results in large deviations from the target geometry,  as can be seen in Figure \ref{fig:urch}, which we attribute to the low fidelity of the linear model. To improve performance, we refine the trajectory generated by the first stage using the high-fidelity ANN model \eqref{eq:nlm}.
In principle, it is possible to skip the first stage and directly embed the ANN model in an optimization problem similar to \eqref{eq:stage1}. However 
numerical experiments suggest that this often leads to poorer performance even compared to the linear model, as the solver tends to converge to poor local minima, or fails to return a feasible solution. Here we take advantage of the output of \eqref{eq:stage1} to fix the time discretization and further improve precision through a second optimization problem based on the nonlinear model. In practice this leads to a more reliable and scalable overall method.

Given the structure of the nonlinear model, which is trained using time series data, we sample the target geometry
at $M$ equally spaced points in time
\begin{equation}
	\bar \Objective_i \doteq  \Objective(s(t_i)): i\in\{0,1, \dots, M\}
\end{equation}
$t_{i+1} = t_i + \DeltatTwo$, where $\DeltatTwo$ is the fixed sample rate of the time series data used to train the nonlinear model.
Given this fixed discretization, the target geometry of the second-stage optimization problem is simply to minimize the deviation from the target trajectory while satisfying the tolerance, speed, and acceleration bounds. This leads to the optimization problem
\begin{mini!}[2]{\ReferenceTwo, \OutputTwo}{\label{eq:stage2-a} \sum_{i=1}^M \|{\OutputTwo_i - \bar\Objective_i}\|^2_2}{\label{eq:stage2}}{}
	\addConstraint{\label{eq:stage2-b} \OutputTwo_{i} = ~}{f_{\mathrm{nlm}}(\ReferenceTwo_{i-h}, \dots, \ReferenceTwo_{i-1}),}{\,i=1,\dots,M}
	\addConstraint{\label{eq:stage2-c} | \OutputTwo_i - \bar\Objective_i | \leq~}{ \Bound_i}{\,i=1,\dots,M}
	\addConstraint{\label{eq:stage2-d} \OutputTwo_i \in ~}{ \mathcal{W}}{\,i=1,\dots,M}
	\addConstraint{\label{eq:stage2-e} | \partial~	\OutputTwo_i | \leq~}{\mathtt{v_{max}}}{\,i=1,\dots,M-1}
	\addConstraint{\label{eq:stage2-f} | \partial^2 \OutputTwo_i | \leq~}{\mathtt{a_{max}}}{\,i=2,\dots,M-1}
	\mathrm{,}
\end{mini!}
where $M$ is the number of discretization points,
$\OutputTwo = \{\OutputTwo_i\}_{i=1}^{M}\subseteq \reals^2$ is the output,
$\ReferenceTwo = \{\ReferenceTwo_i\}_{i=1}^{M}\subseteq \reals^2$ is the input trajectory,
$\bar \Xi=\{\bar \Xi_i\}_{i=1}^{M}\subseteq \reals^2$ is the sampled target geometry,
$f_{\mathrm{nlm}}$ is the nonlinear ANN model \eqref{eq:nlm},
$\mu_i$ is the tolerance bound evaluated for each discretization point,
$\mathcal{W}$ are the limits of the working space of the device,
$\mathtt{v_{max}}$ is the maximum allowable speed, and
$\mathtt{a_{max}}$ is the acceleration limit.

The cost \eqref{eq:stage2-a} penalizes deviations of the output from the target geometry. 
The constraint \eqref{eq:stage2-b} imposes the nonlinear system dynamics modeled by the ANN,
\eqref{eq:stage2-c} constrains the output to be within the tolerance bound for each discretization point,
While \eqref{eq:stage2-d}, \eqref{eq:stage2-e} and \eqref{eq:stage2-f} ensure that the output, its velocity and acceleration stay within working space and operational limits of the device.

	In practice, to ensure feasibility, the tolerance constraint is replaced with an exact $L_1$ penalty which is implemented using slack variables~\cite{goodwin2006constrained}. For longer trajectories, due to memory constraints, the optimization problem can be solved with a receding horizon strategy.
	In this case we use an horizon of $11$ steps.
	The maximum acceleration limit $\mathtt{a_{max}}$ is the main parameter determining the total trajectory time, tuning the solution along the accuracy/trajectory time trade-off curve.

\subsection{Implementation Details}
\if 01
\begin{itemize}
	\item Julia, Casadi, Ipopt etc.
	\item Solution times etc.
	\item Scaling results etc. can go here
\end{itemize}
\fi

The optimization problems defined in \eqref{eq:stage1} and \eqref{eq:stage2} are both nonlinear and are solved using \texttt{IPOPT} \cite{Ipopt}, with the JuMP \cite{JuMP}, and Casadi \cite{Casadi} interfaces in Julia and Python respectively. 
The first stage optimization problem \eqref{eq:stage1} is initialized by taking as $\ReferenceOne$ the target geometry traversed at a constant speed.
The solution of \eqref{eq:stage1} is then used to initialize the second-stage \eqref{eq:stage2}. 

The computer used throughout this paper runs Arch Linux, with Linux kernel version 5.15, and it is equipped with a \verb|GeForce RTX 2080 Ti| GPU with $12~\mathrm{GB}$ of dedicated memory, an \verb|Intel(R) Core(TM) i9-9900K| CPU $@~3.60~\mathrm{GHz}$ and $64~\mathrm{GB}$ of RAM.

For the problem sizes solved in this paper, the first-stage optimization takes approximately 1 minute to complete ($N \approx 10^3$), and the second-stage takes approximately between 1 to 4 hours depending on the trajectory time ($M \in \{200, \dots, 1000\}$). Lower $\mathtt{a_{max}}$ results in slower trajectories which take more time to compute due to the increased number of function evaluations required. 
Training of the ANN model takes approximately $24~\mathrm{h}$ per axis.

\section{Experimental Results} \label{sec:results}

In this section, we apply our proposed data-driven optimization methodology from Section~\ref{sec:approach} to the experimental apparatus in~\ref{subsec:setup}, using the test cases in~\ref{testcases}.
Our experiments demonstrate that the methodology is capable of improving system performance relative to the baseline -- the trajectory obtained with the low-level controller following a not optimized input trajectory -- by tracing desired geometries faster and more precisely.

\subsection{Individual trajectories}
\label{ssec:Individual}

We first focus on the letter ``r'' as a case study and consider a scenario with an acceleration limit of $1~\mathrm{m\,s^{-2}}$; the individual results for the other letters in the test case are comparable and will be discussed collectively in Section \ref{sec:trade-off}.
Applying our method results in an optimized input trajectory with a total time of $T=0.807~\mathrm{s}$.
The experimental result of this new input trajectory is compared with a baseline performed in the same total time.
For this the original, non-optimized shape (in this case the letter ``r'') is sampled at a constant progress speed
$	\ReferenceTwo(t) = \Objective\left(t \frac{S}{T}\right)\mathrm{,}$
where $S$ is the total path length, which for a closed shape corresponds to the perimeter.
\begin{figure}[t]
	\begin{center}
		\includegraphics[width=.6\columnwidth]{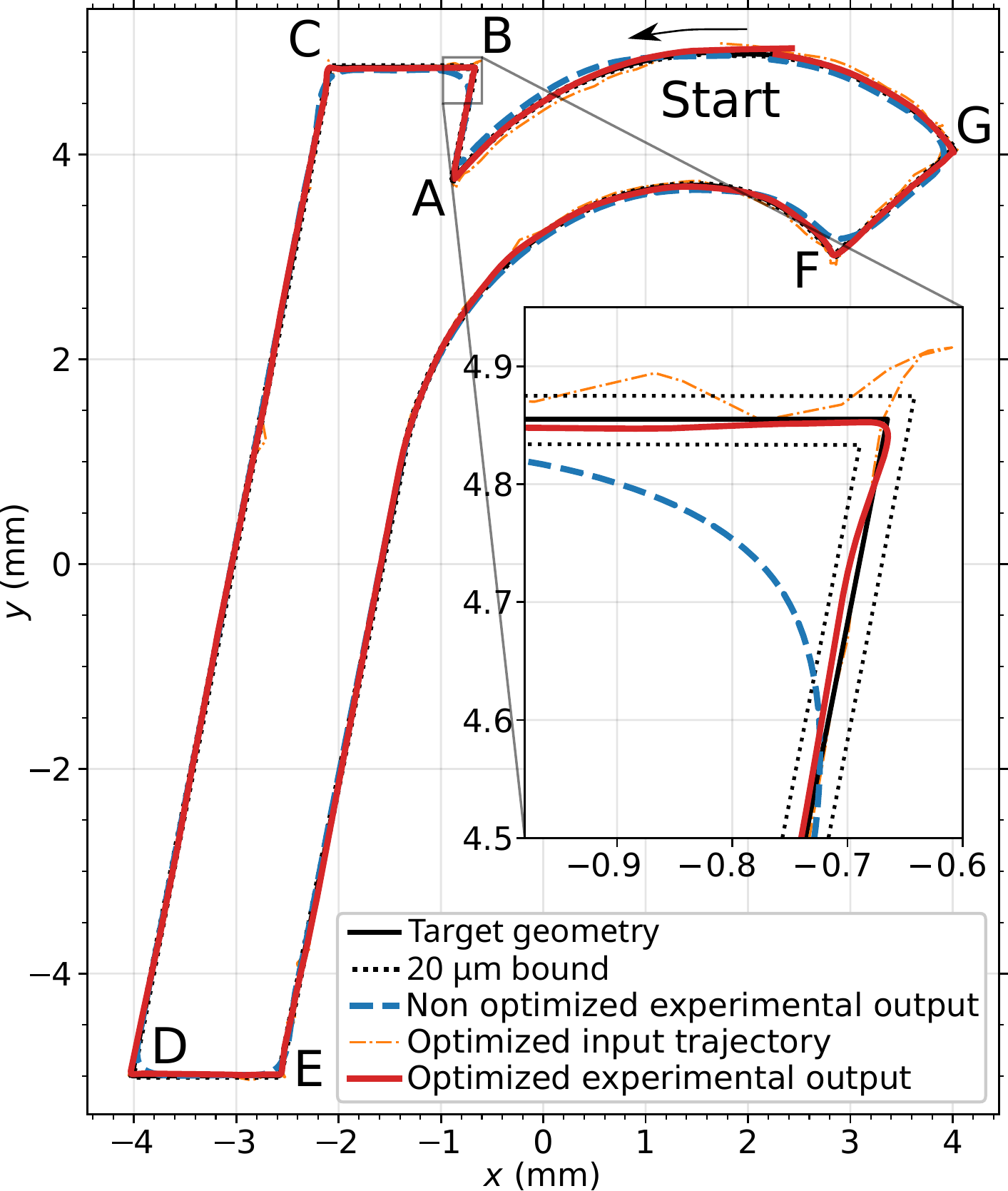} 
		\caption{Letter ``r'' of the ETH Z\"{u}rich Logo. Optimized with $\mathtt{a_{max}} = 1~\mathrm{m\,s^{-2}}$, including a detail view of station B. Note how the optimized input trajectory significantly deviates from the target geometry near the corner in B to compensate for the dynamics of the machine. This leads to significantly smaller error between the output and the target geometry compared to using the state-of-practice (non-optimized) input trajectory.}
		\label{fig:r-xy-all}
		\label{fig:r-xy-detail}
	\end{center}
\end{figure}
\if 01
\begin{figure}[t]
	\begin{center}
		\includegraphics[width=\resultsscaling\columnwidth]{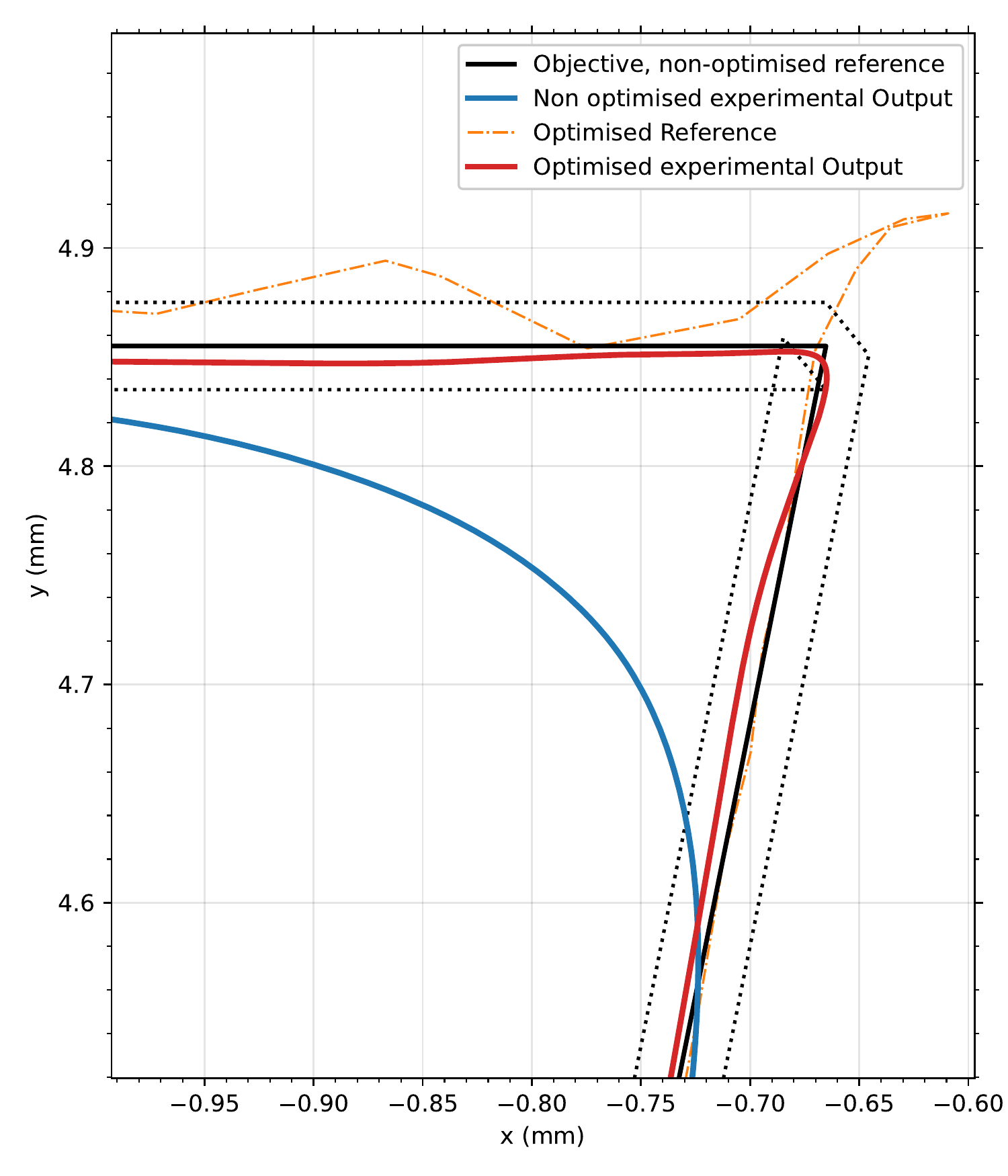} 
		\caption{A close-up of station B on the letter ``r''. The optimized input trajectory overshoots the geometry to compensate for the dynamics of the machine, leading to significantly less deviation. $\mathtt{a_{max}}= 1~\mathrm{m\,s^{-2}}$}
	\end{center}
\end{figure}
\fi
\begin{figure}[t]
	\begin{center}
		\includegraphics[width=\resultsscaling\columnwidth]{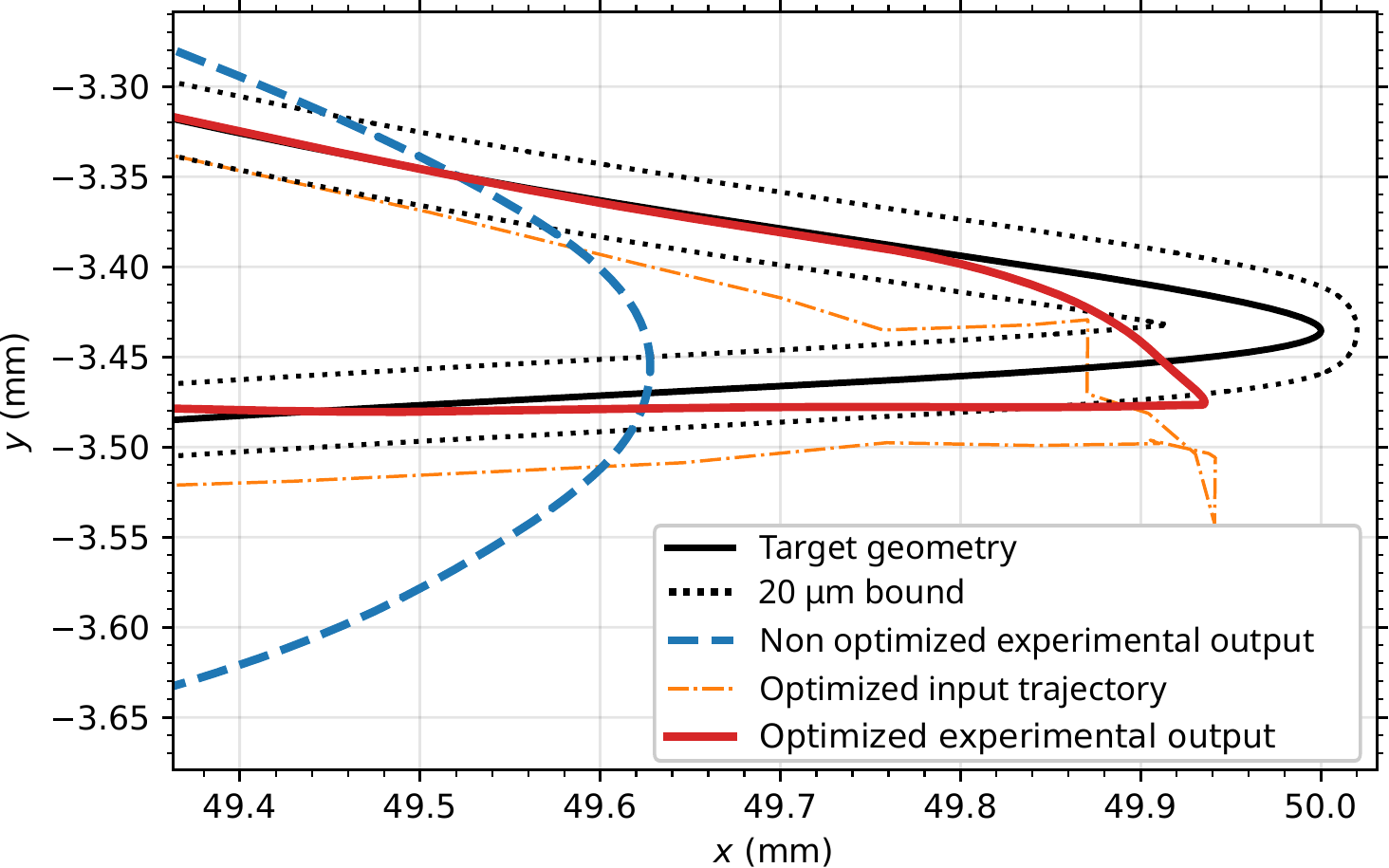} 
		\caption{A close-up of the trailing edge of the airfoil test case with $\mathtt{a_{max}} = 2~\mathrm{m\,s^{-2}}$. The optimized trajectory is able to track the desired geometry more precisely than the baseline, though still not within the $20~\mathrm{\mu m}$ tolerance band. The optimized input trajectory is not constrained by the tolerance band.}
		\label{fig:o-xy-detail}
	\end{center}
\end{figure}

\begin{figure*}
	\begin{subfigure}{\columnwidth}
		\begin{center}
		\includegraphics[width=\resultsscaling\columnwidth]{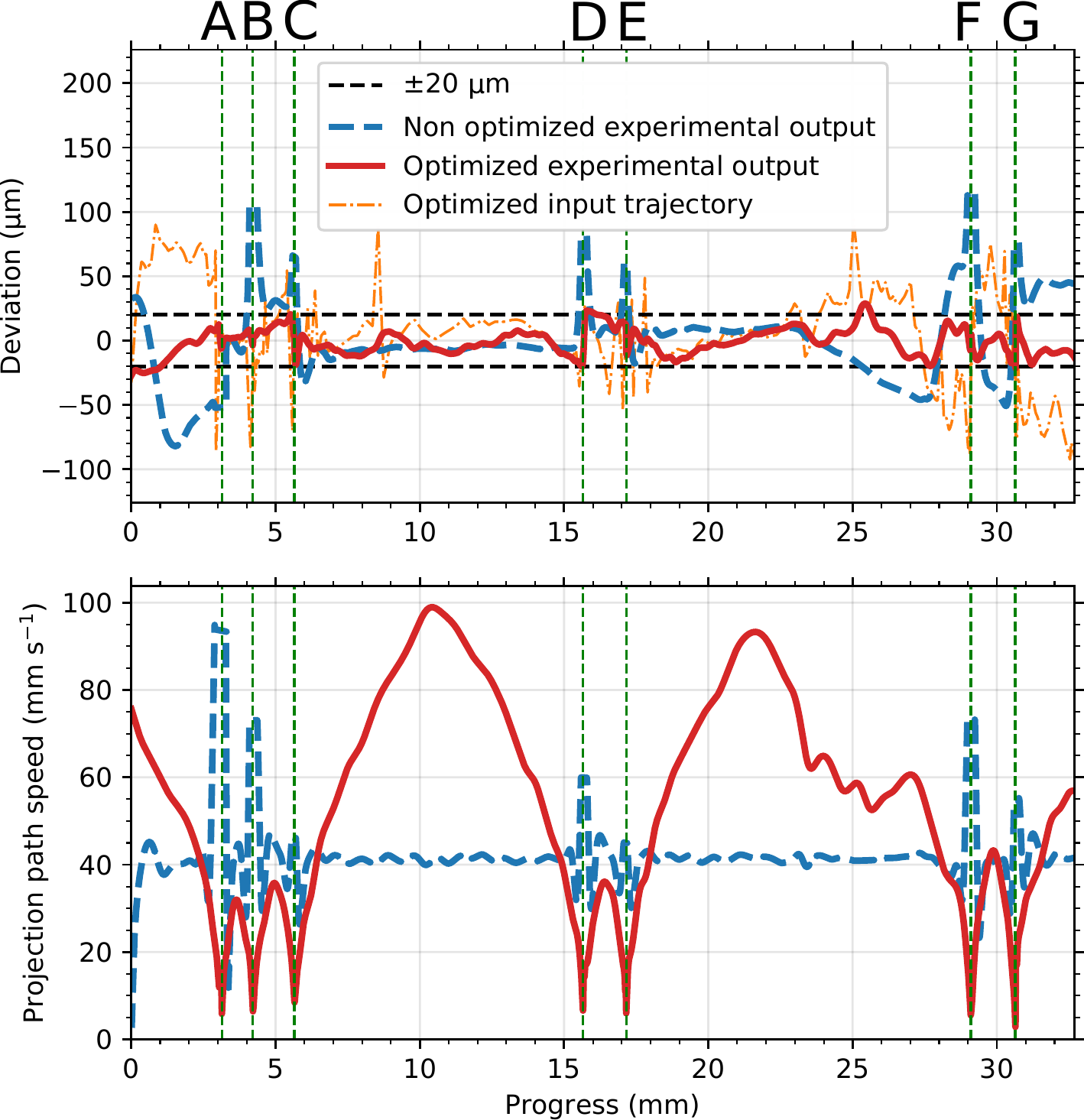} 
		\caption{Letter ``r'' of the ETH Z\"{u}rich logo optimized with $\mathtt{a_{max}} = 1~\mathrm{m\,s^{-2}}$. The letters A-G refer to the points labeled in Figure \ref{fig:r-xy-detail}.\label{fig:r-progress}}
		\end{center}
	\end{subfigure}\hfill
	\begin{subfigure}{\columnwidth}
		\begin{center}
		\includegraphics[width=\resultsscaling\columnwidth]{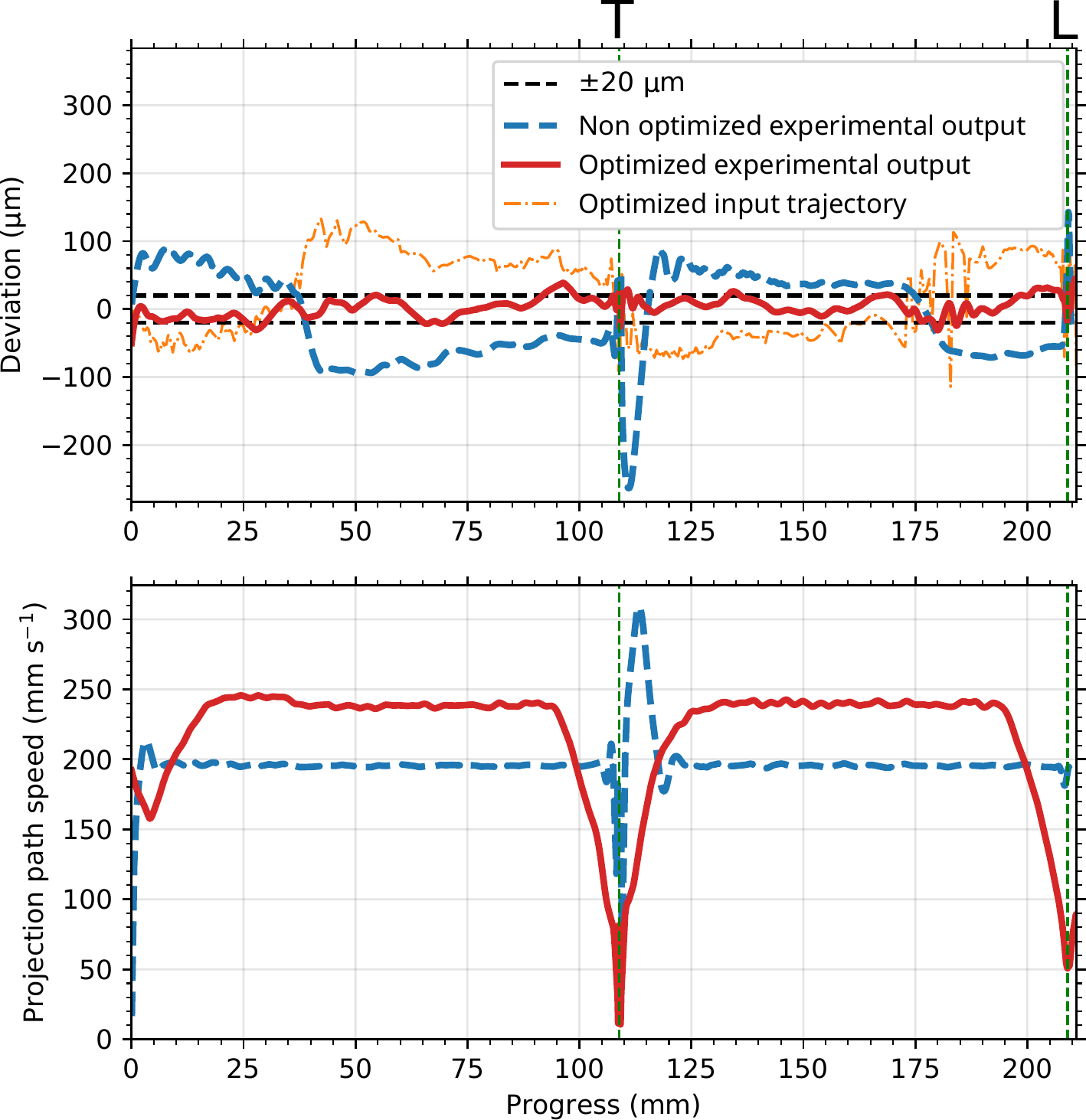} 
		\caption{Airfoil test case, optimized with $\mathtt{a_{max}} = 2~\mathrm{m\,s^{-2}}$. The letters L and T refer to the leading and trailing edges labeled in Figure~\ref{fig:airfoil}.\label{fig:o-progress}}
		\end{center}
	\end{subfigure}
	\caption{Top panels show the deviation from the target geometry for the optimized and non-optimized output, as well as the optimized input trajectory. The non-optimized input trajectory by construction does not deviate from the target geometry. Bottom panels shows the velocity of the optimized and non-optimized outputs projected onto the target geometry.}
	\label{fig:progess}
\end{figure*}

Figure~\ref{fig:r-xy-all} compares the output and input trajectories for the optimized and baseline cases in the $x-y$ plane while Figure~\ref{fig:r-progress} displays the same data as a function of progress.

Analyzing Figures~\ref{fig:r-xy-all} and \ref{fig:r-progress}, we see that the optimized output effectively stays within the selected $\Bound = 20~\mathrm{\mu m}$ while the baseline output deviates by more than $100 \mu m$. The optimized output trajectory speeds up in areas with low curvature and slows down near corners or other intricate features. 
%
Figure~\ref{fig:r-progress} also illustrates that the optimized input trajectory deviates aggressively from the target geometry near areas where the baseline output performs poorly. This can be interpreted as an attempt to ``cancel out'' the error; The detail in Figure~\ref{fig:r-xy-detail} illustrates this behavior in the $x-y$ plane.

Figure~\ref{fig:o-progress} display the same information for the airfoil test geometry with a maximum acceleration of $2~\mathrm{m\,s^{-2}}$. The baseline and optimized trajectories complete the part in $1.081~\mathrm{s}$. Similarly to the letter ``r'' example, the optimized trajectory slows down near intricate features, in this case the leading and trailing edges of the airfoil, and accelerates between them. Figures~\ref{fig:o-xy-detail} and \ref{fig:o-progress} show that the optimized input again seeks to ``compensate'' for deviations in the baseline trajectory.

\if01
\begin{figure}[t]
	\begin{center}
		\includegraphics[width=\resultsscaling\columnwidth]{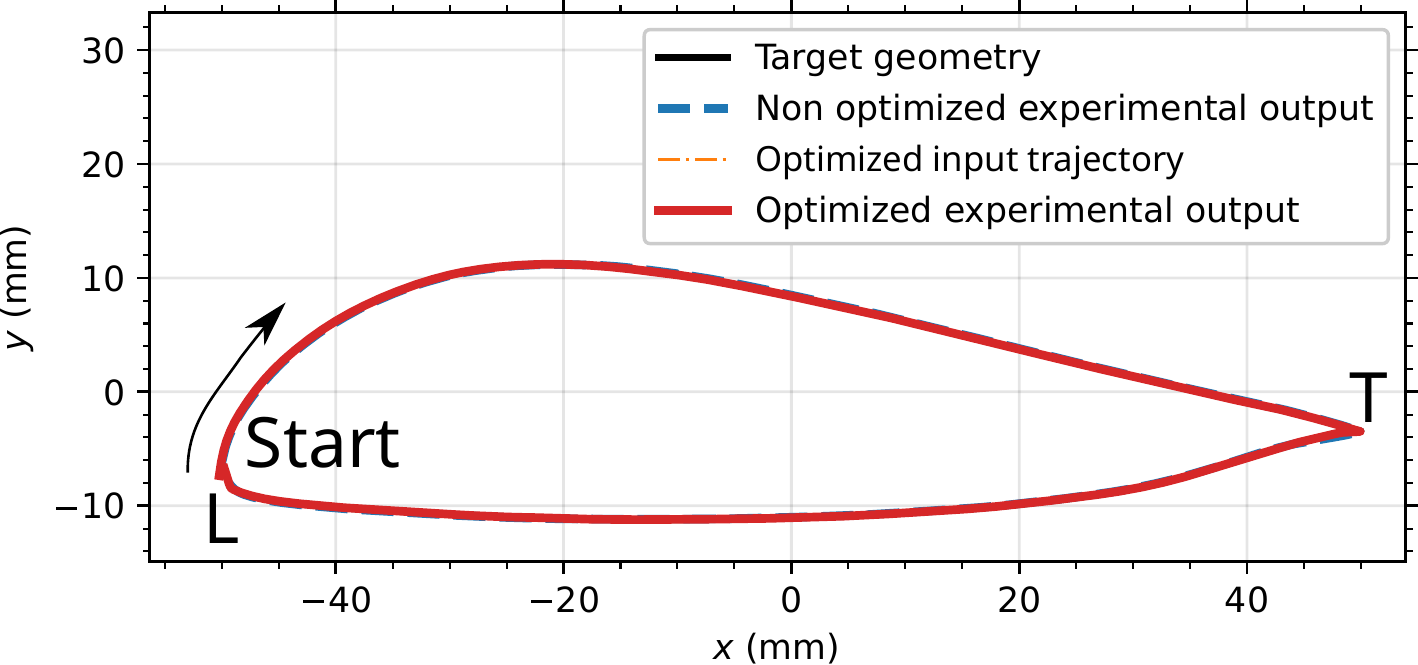} 
		\caption{Experimental results for the \Airfoil~test case optimized with $\mathtt{a_{max}} = 2~\mathrm{m\,s^{-2}}$.
		L denotes the leading and T the trailing edge of the airfoil. The shape is traversed in the clockwise direction starting from the leading edge.}
		\label{fig:o-xy-all}
	\end{center}
\end{figure}
\fi

To summarize, we observe that the optimizer exploits two main mechanisms to improve performance of the machine relative to the baseline system with the same trajectory time:
\begin{enumerate*}
\item It reduces the speed of the input trajectory near intricate geometric features and increases it in areas with low curvature. 
\item It ``compensates'' for errors in the baseline trajectory by moving the optimized input trajectory in an opposing direction.
\end{enumerate*}
In effect, the optimizer exploits information encapsulated in the data-driven models to push the limits of system performance and adapt the input trajectory to the machine capabilities, for the selected maximum acceleration value.

\subsection{Precision-accuracy Trade-off}\label{sec:trade-off}
In precision motion systems there is a fundamental trade-off between speed and accuracy caused primarily by the inertia of the machine. In this section, we demonstrate that our proposed methodology is able to improve system performance by shifting this trade-off.

In our formulation \eqref{eq:stage1}-\eqref{eq:stage2}, the trade-off between speed and accuracy is controlled by the parameter $\mathtt{a_{max}}$ which limits the maximum acceleration of the input trajectory. We generated trade-off curves for both the letters and airfoil test geometries by conducting experiments for different values of this parameter between $0.1~\mathrm{m\,s^{-2}}$ and $3.3~\mathrm{m\,s^{-2}}$ for the letters and between $0.1~\mathrm{m\,s^{-2}}$ and $10~\mathrm{m\,s^{-2}}$ for the airfoil, and tested each resulting input trajectory experimentally.
Similarly to the two individual test cases presented above, the non-optimized input trajectories are subsequently run with the same total time as the one found for the optimized input trajectories.

We use the normalized $L_1= \frac{1}{M} \sum_{i=1}^M  \min_s \left|\left|\OutputTwo_{i} - \Objective(s)\right|\right|_2$ and $L_2= \sqrt{ \frac{1}{M} \sum_{i=1}^M  \min_s \left|\left|\OutputTwo_{i} - \Objective(s)\right|\right|_2^2}$ (rms) norms as well as the $L_\infty=\max_{i\in\{1,\dots,M\}} { \min_s \left|\left|\OutputTwo_{i} - \Objective(s)\right|\right|_2}$ norm to quantify the deviation between the output $\OutputTwo$ and target geometry $\Objective$, both in simulation and experimentally.
\if01
\begin{equation}
	{L_1}= \frac{1}{M} \sum_{i=1}^M  \min_s \left|\left|\OutputTwo_{i} - \Objective(s)\right|\right|_2,
\end{equation}

\begin{equation}
	{L_2} = \sqrt{ \frac{1}{M} \sum_{i=1}^M  \min_s \left|\left|\OutputTwo_{i} - \Objective(s)\right|\right|_2^2},
\end{equation}

\begin{equation}
	{L_\infty} = \max_{i\in\{1,\dots,M\}} { \min_s \left|\left|\OutputTwo_{i} - \Objective(s)\right|\right|_2}.
\end{equation}
\fi

\if 01
two vectors $a,b\in\reals^{M\times2}$:
\begin{equation}
	{L_1}(\OutputTwo) = \frac{1}{M} \sum_{i=1}^M |\OutputTwo_i - \bar\Objective_i|,
\end{equation}

\begin{equation}
	{L_2}(\OutputTwo) = \sqrt{ \frac{1}{M} \sum_{i=1}^M (\OutputTwo_i - \bar\Objective_i)^2},
\end{equation}

\begin{equation}
	{L_\infty}(\OutputTwo) = \max_{i\in\{1,\dots,M\}} { |\OutputTwo_i - \bar\Objective_i|},
\end{equation}
\begin{equation}
	{L_1}(a, b) = \frac{1}{M} \sum_{i=1}^M  \sqrt{\sum_{j=1}^2 (a_{ij} - b_{ij})^2},
\end{equation}

\begin{equation}
	{L_2}(a, b) = \sqrt{ \frac{1}{M} \sum_{i=1}^M \sum_{j=1}^2 (a_{ij} - b_{ij})^2},
\end{equation}

\begin{equation}
	{L_\infty}(a, b) = \max_{i\in\{1,\dots,M\}} {\sqrt{\sum_{j=1}^2 (a_{ij} - b_{ij})^2}}.
\end{equation}
\fi

\begin{figure}[t]
	\begin{center}
		\includegraphics[width=\resultsscaling\columnwidth]{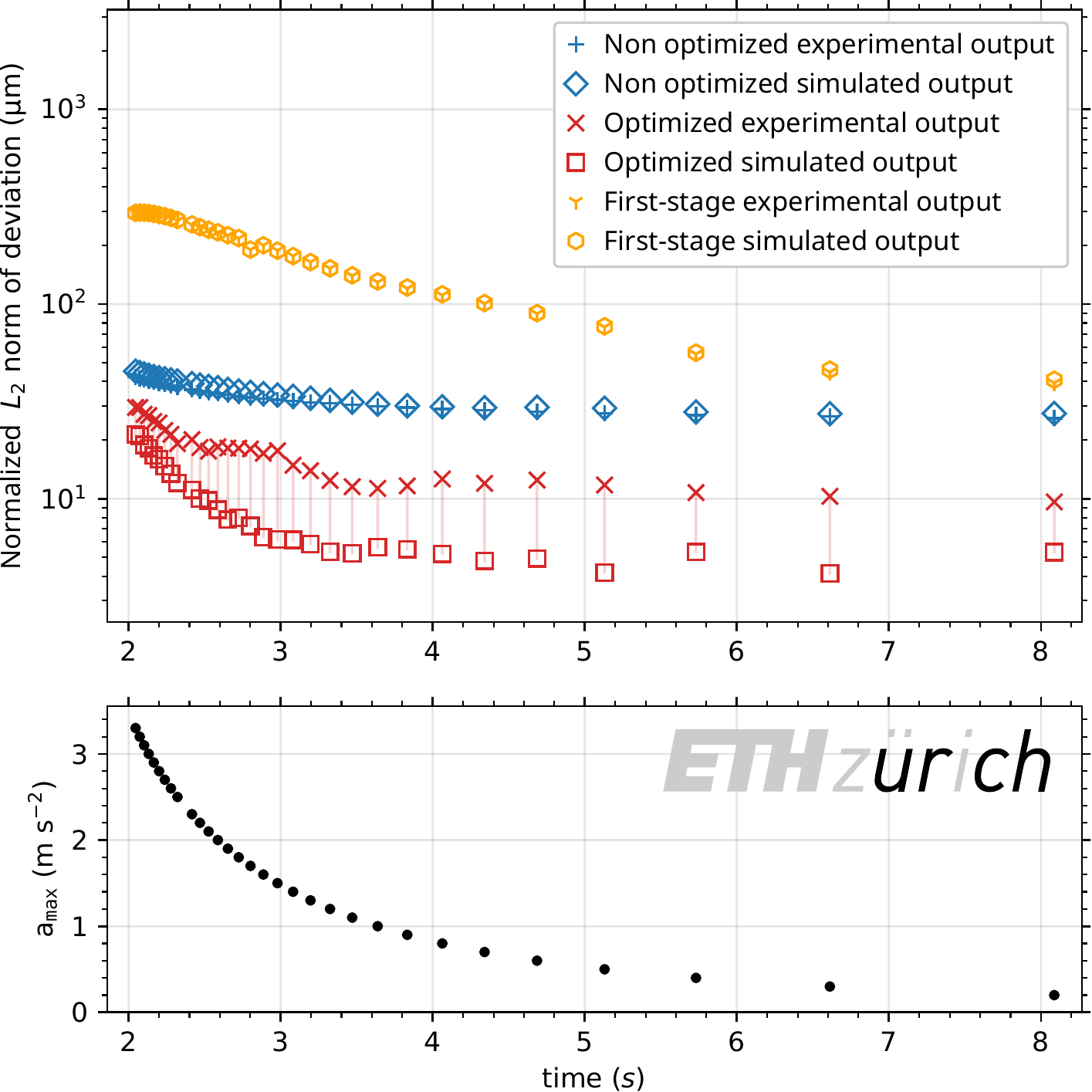} 
		\caption{Precision-speed trade-off curve for the letters test case.
		The top panel shows the normalized $L_2$ deviation of the outputs from the target geometry. 
		The bottom panel shows the total time to perform the trajectory and the $\mathtt{a_{max}}$ used in optimization. 
		To determine the performance at a given acceleration value, one can use the lower panel to determine the time needed to complete the trajectory for a given acceleration, then retrieve the deviation for the corresponding time from the top plot.}
		\label{fig:urch}
	\end{center}
\end{figure}

The results are shown in Figure~\ref{fig:urch}. In all cases the optimized experimental output results in significantly more precise trajectories for a given part completion time than the non-optimized trajectories.
The input trajectories optimized in the first stage exhibit higher deviations than the non-optimized trajectories. This is to be expected given the prediction error of the linear model, as quantified in Table \ref{tab:linear-model-acc}.
Moreover, the simulated deviation is lower than the experimental one due to model mismatch.
%
%
%
%
Figure~\ref{fig:urch} also show that for the same deviation values the optimized trajectories take less time to complete. For example, in the letters test case the non-optimized version takes $8.1~\mathrm{s}$ and achieves a deviation of $26~\mathrm{\mu m}$, while the optimized version takes only $2.2~\mathrm{s}$ for an identical deviation. In this case the optimized trajectory reduces the time needed to trace the shape in 73\%. For the airfoil test case a similar analysis yields a reduction of 57\% for a deviation of $47~\mathrm{\mu m}$.
In Table \ref{tab:perimprov} we show the results for $5$ different test cases, each at two different $\mathtt{a_{max}}$ scenarios. For all cases our method improves system performance in $L_1$, $L_2$ and $L_\infty$ norms.

\if01
\begin{figure}[t]
	\begin{center}
		\includegraphics[width=\resultsscaling\columnwidth]{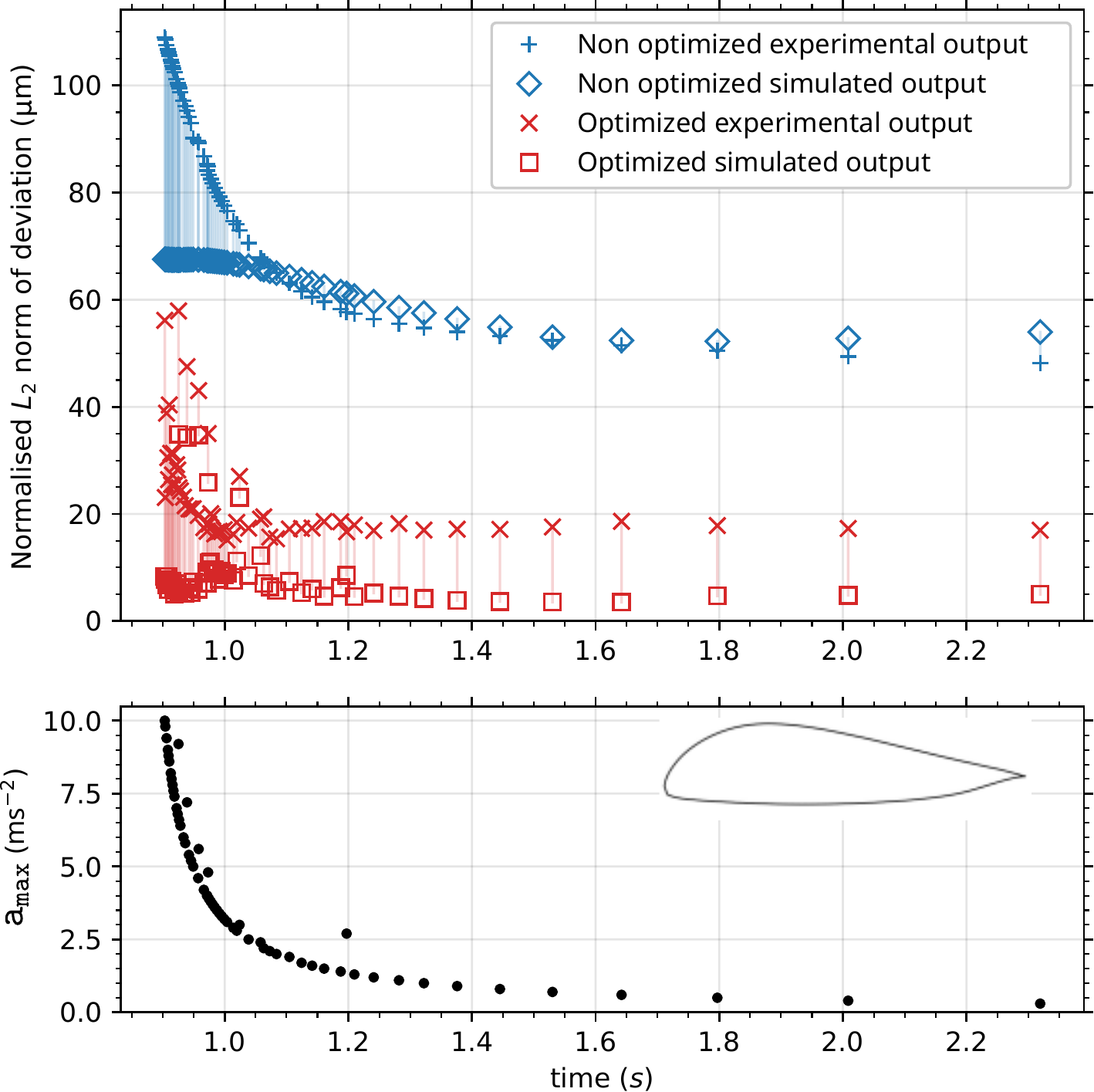} 
		\caption{Precision-speed trade-off curve for the airfoil test case. The interpretation of the plots is similar to Figure~\ref{fig:urch}
		}
		\label{fig:o-devL2}
	\end{center}
\end{figure}
\fi
\begingroup
\renewcommand{\arraystretch}{1.3}
\begin{table}
	\begin{center}
		\begin{tabular}{@{}*{9}{r}@{}}
			\toprule
			&\multicolumn{4}{c}{$\mathtt{a_{max}} = 1.0~ \mathrm{m\,s^{-2}}$} &\multicolumn{4}{c}{$\mathtt{a_{max}} = 3.0~ \mathrm{m\,s^{-2}}$} \\
			\cmidrule(lr){2-5} \cmidrule(lr){6-9}
Shape	&time			&$L_1$	&$L_2$	&$L_\infty$ &time			&$L_1$	&$L_2$	&$L_\infty$	\\
--		&$\mathrm{s}$	&\%		&\%		&\%			&$\mathrm{s}$	&\%		&\%		&\%	\\
\midrule
Letter u	&1.052	&52.7	&60.0	&63.0 	&0.618	&36.4	&38.7	&46.1\\
Letter r	&0.807	&58.9	&64.3	&75.1 	&0.473	&32.8	&29.7	&42.9\\
Letter c	&0.807	&80.0	&78.5	&74.4 	&0.473	&40.2	&38.6	&54.3\\
Letter h	&0.971	&26.6	&32.8	&31.0 	&0.568	&22.6	&27.0	&43.1\\ 
\Airfoil	&1.322	&73.5	&69.0	&64.9 	&1.024	&75.0	&63.0	&28.1\\
				\bottomrule
		\end{tabular}
\caption{\label{tab:perimprov} Percent improvement in accuracy after applying the proposed input trajectory design method compared to the default control without input trajectory optimization. Experimental results for different shapes and $\mathtt{a_{max}}$. }
	\end{center}
\end{table}
\endgroup

\subsection{Effect of horizon length}

\begin{figure}[t]
	\begin{center}
		\includegraphics[width=.7\columnwidth]{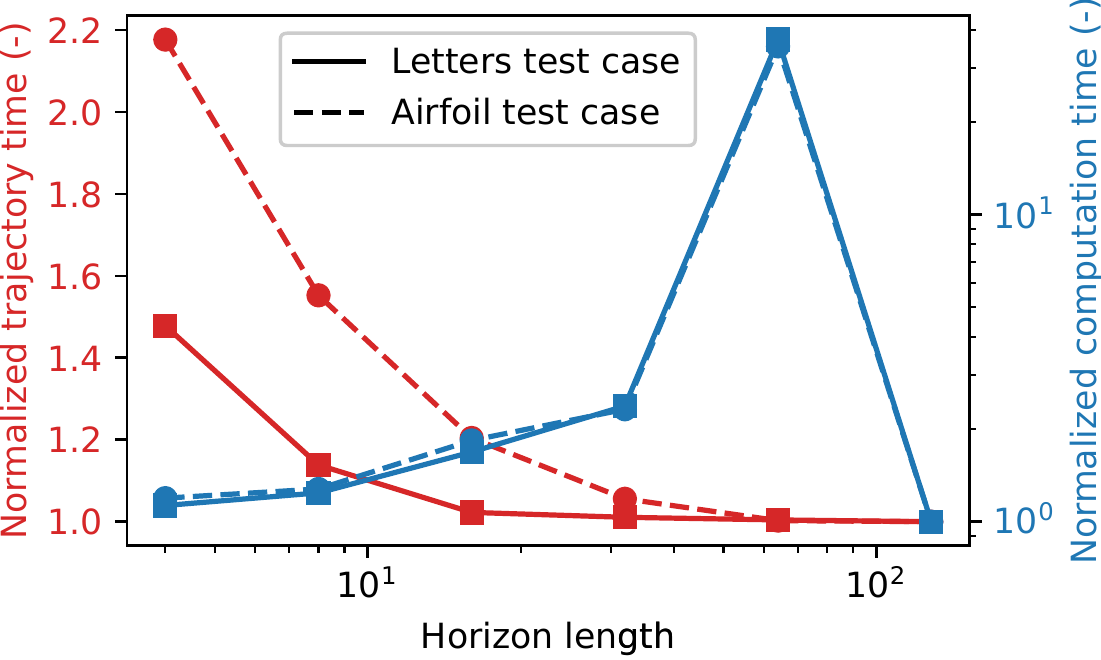} 
		\caption{Trajectory time and computation time of the first stage as function of the horizon length using a receding horizon strategy. Both times are normalized by the one-shot optimization case, the right-most points where the horizon is equal to the number of points used in the discretization $N=128$. In this study the tolerance band is set to $200\,\mathrm{\mu m}$.}
		\label{fig:time_vs_horizon}
	\end{center}
\end{figure}
In Figure \ref{fig:time_vs_horizon} we study the effect of optimizing the reference trajectory with a receding horizon strategy on the first stage optimization.
We observe that the trajectory time is the smallest for the one-shot optimization, and increases with a reduction of the horizon length. The increase is more pronounced for the Airfoil test case, which can be traced at higher speeds compared to the Letters test case. 
The computation time is also the smallest for the one-shot optimization. Short horizons are fast to compute, but require the largest number of optimization problems to be solved. The maximum computation time is reached for long horizons, where each optimization problem takes some time to compute and many are required.


\section{Conclusion} \label{sec:conclusion}
In this paper we proposed a method that improves the precision vs. productivity trade-off of a PMS. We use models built exclusively from experimental data, and only modify the input trajectory provided to the closed loop control system.
Experimental data obtained for shapes outside of the training dataset corroborates simulation results and shows that the method can significantly improve system performance, 
reliably shifting the precision vs. productivity trade-off curve across a wide range of operating conditions.
This is accomplished by exploiting variable speeds (possible since the full trajectory is optimized in one go), and error compensation while respecting system operational constraints.

Future work will focus on reducing the computational load of the offline input trajectory optimization and further increasing the ANN complexity to improve the prediction accuracy. In another direction, the ANN can be combined with an ILC loop. In an ILC setting the number of trials can be reduced by using the ANN gradient information, either independently or with an initial solution provided by the method we proposed here, taking advantage of the strengths of both methods.

\section*{Acknowledgment}
The authors would like to thank Dr. Natanael Lanz for his support with the operation of the experimental apparatus.

\bibliography{database.bib}
\begin{IEEEbiography}[{\includegraphics[width=1in,height=1.25in,clip,keepaspectratio]{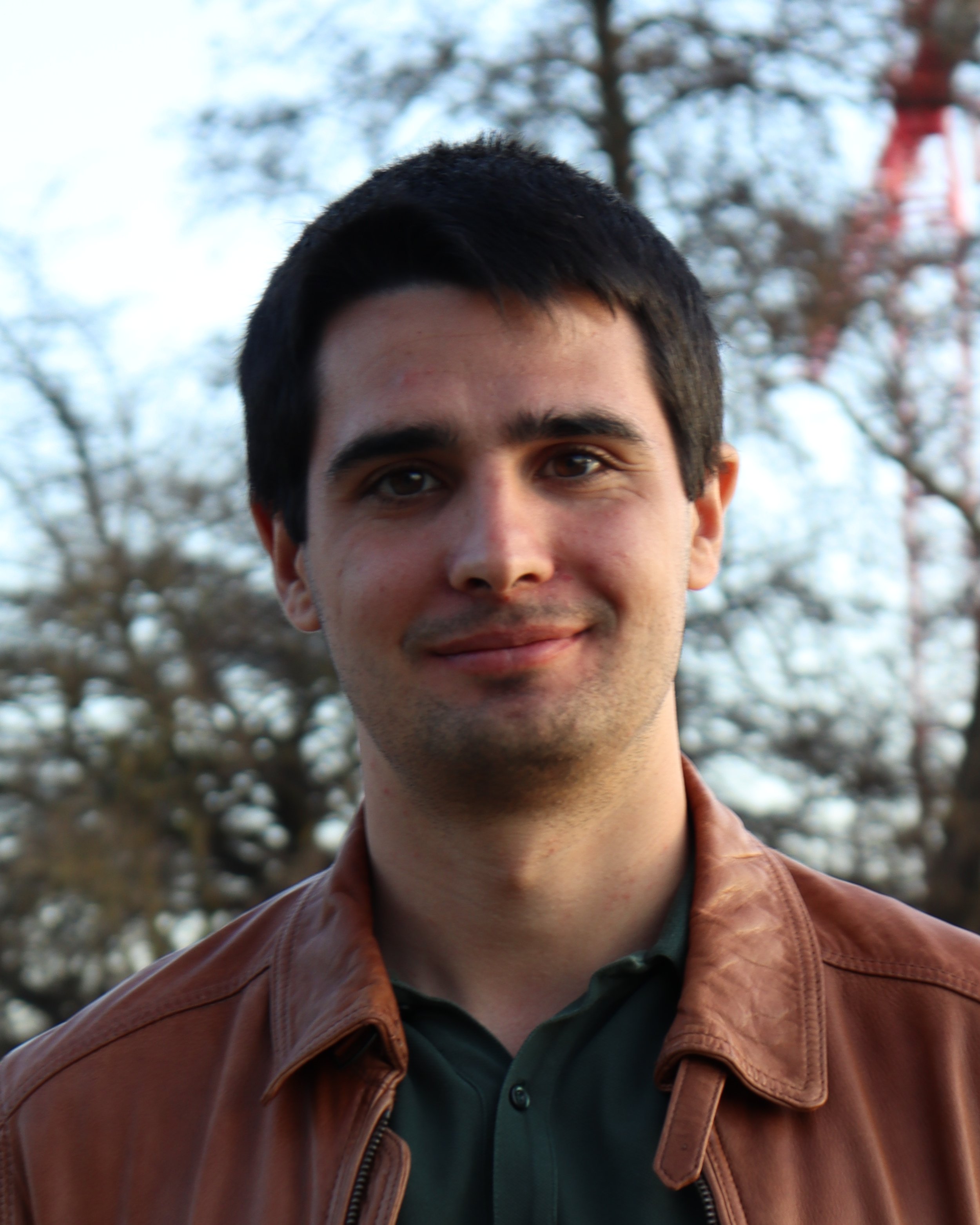}}]{Samuel Balula} 
	received his B.Sc. and M.Sc. in Engineering Physics from Instituto Superior Técnico, University of Lisbon, Portugal in 2014 and 2016 respectively.
Between 2017-2018 he was head of research at a tech startup.
He is currently a PhD student in the Automatic Control Laboratory at ETH Zürich, Switzerland.
His research interest include optimal non-linear control, trajectory planning, data-driven control and optimal experimental design with applications in robotics and manufacturing.

\end{IEEEbiography}

\begin{IEEEbiography}[{\includegraphics[width=1in,height=1.25in,clip,keepaspectratio]{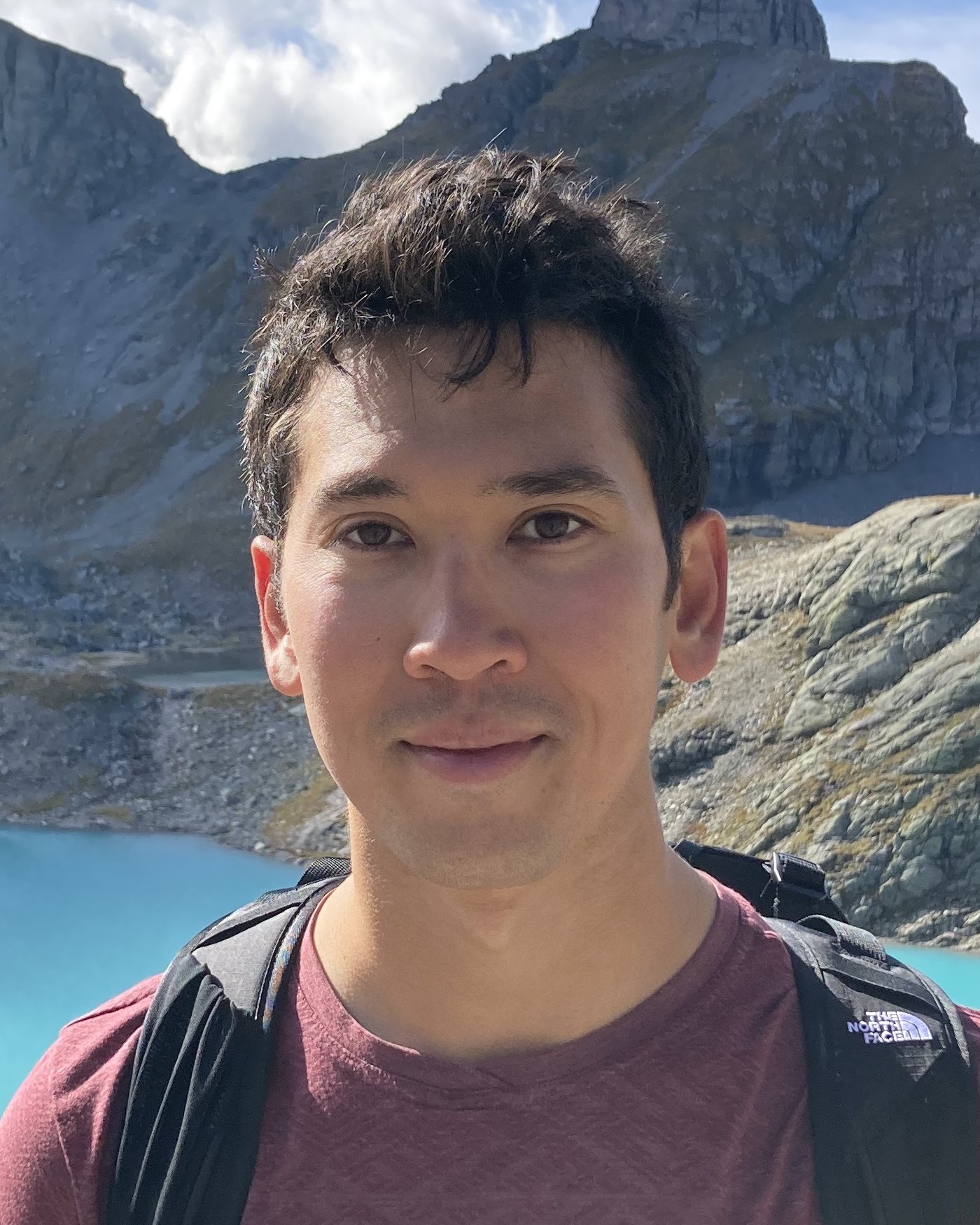}}]{Dominic Liao-McPherson} 
	received his BASc. in Engineering Science from the University of Toronto in 2015 and his PhD. in Aerospace Engineering and Scientific Computing from the University of Michigan, Ann Arbor, in 2020. He is currently a postdoctoral researcher at the ETH Zürich Automatic Control Laboratory. His research interests include real-time optimization, game-theoretic multi-agent control and numerical algorithms with applications in robotics, energy, and manufacturing.

\end{IEEEbiography}

\begin{IEEEbiography}[{\includegraphics[width=1in,height=1.25in,clip,keepaspectratio]{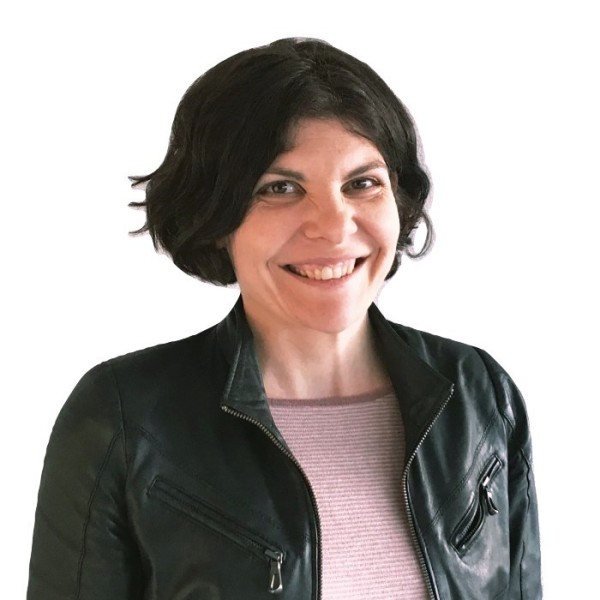}}]{Alisa Rupenyan} 
     received the BSc in Engineering physics and the MSc degree in Laser physics from the University of Sofia, Sofia, Bulgaria, in 2004 and 2005, and the Ph.D. degree from the Department of Physics and Astronomy of Vrije Universiteit Amsterdam, Amsterdam, The Netherlands in 2010. Between 2011-2014 she was a postdoctoral fellow at ETH Z{\"u}rich, Switzerland, and between 2014-2018 a lead scientist in a robotic start-up. She is currently a group leader in automation at inspire, the technology transfer unit at ETH Z{\"u}rich, and a senior scientist and PI at the Automatic Control Laboratory there. Her research interests are on the intersection between machine learning, control, and optimization for industrial applications and robotics.
\end{IEEEbiography}

\begin{IEEEbiography}[{\includegraphics[width=1in,height=1.25in,clip,keepaspectratio]{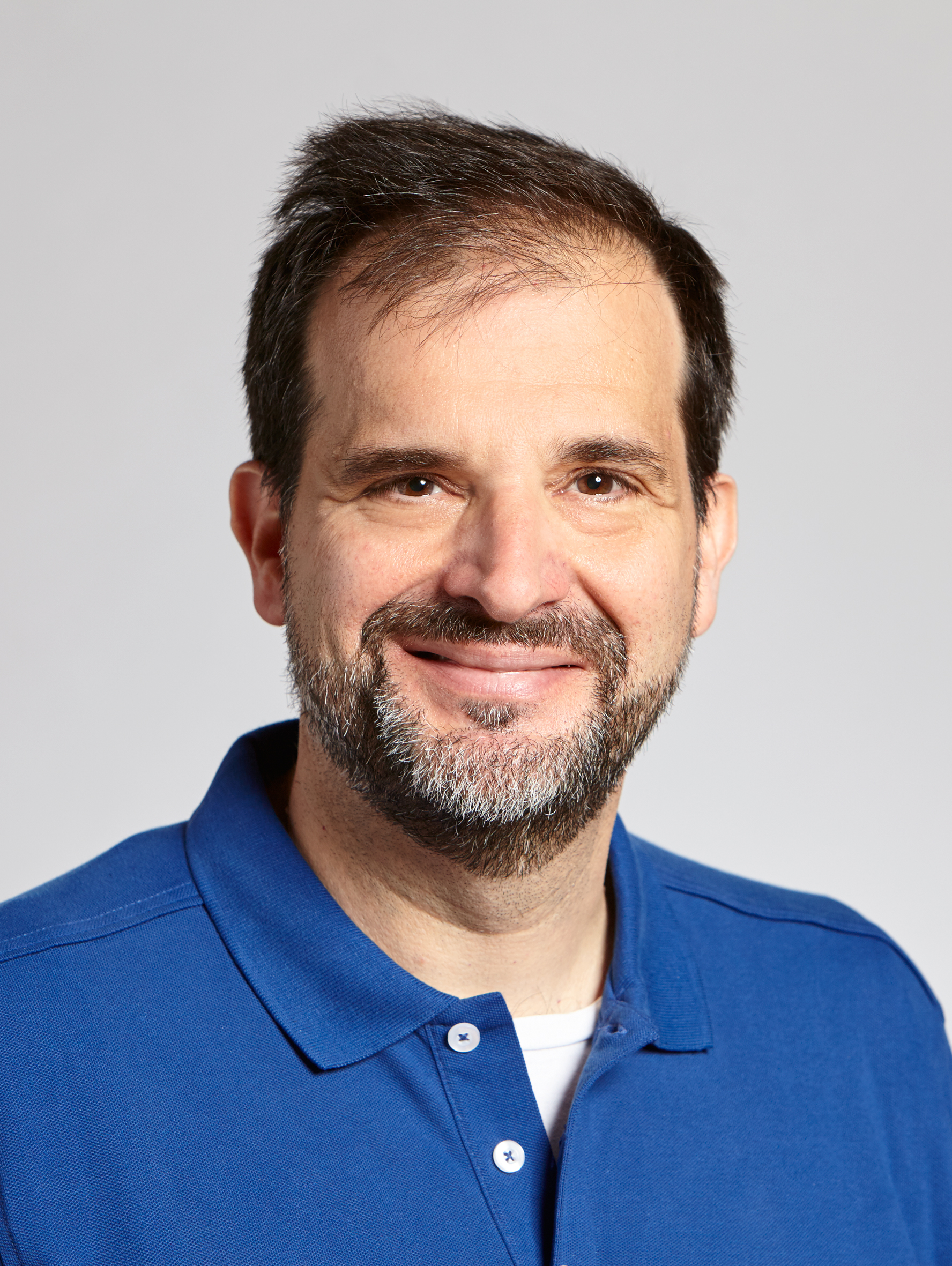}}]{John Lygeros} 
	received a B.Eng. degree in 1990 and an M.Sc. degree in 1991 from Imperial College, London, U.K. and a Ph.D. degree in 1996 at the University of California, Berkeley. After research appointments at M.I.T., U.C. Berkeley and SRI International, he joined the University of Cambridge in 2000 as a University Lecturer. Between March 2003 and July 2006 he was an Assistant Professor at the Department of Electrical and Computer Engineering, University of Patras, Greece. In July 2006 he joined the Automatic Control Laboratory at ETH Zurich where he is currently serving as the Professor for Computation and Control and the Head of the laboratory. His research interests include modelling, analysis, and control of large scale systems, with applications to biochemical networks, energy systems, transportation, and industrial processes. John Lygeros is a Fellow of IEEE, and a member of IET and the Technical Chamber of Greece. Since 2013 he is serving as the Vice-President Finances and a Council Member of the International Federation of Automatic Control and since 2020 as the Director of the National Center of Competence in Research “Dependable Ubiquitous Automation” (NCCR Automation).

\end{IEEEbiography}
\vfill

\end{document}